%% file: Nonlinear_Field.tex
\documentclass[aps,prd,eqsecnum,nofootinbib,superscriptaddress,floats,floatfix,tightenlines]{revtex4}

\usepackage{graphicx}
\usepackage{indentfirst}
\usepackage{latexsym}
\usepackage{multirow}
\usepackage{tabls}
\usepackage{color}
\usepackage{subfigure}

\usepackage{amsfonts}
\usepackage{amsmath}
\usepackage{amssymb}
\usepackage{bm}
\usepackage{mathrsfs}

\usepackage{feyn}

\def\b{ \beta }

\def\cR{ {\cal R} }

\def\b0{ {\bf 0} }

\def\lsim{\mathrel{\rlap{\lower3pt\hbox{\hskip1pt$\sim$}}
    \raise1pt\hbox{$<$}}}                % less than or approx. symbol
\def\gsim{\mathrel{\rlap{\lower3pt\hbox{\hskip1pt$\sim$}}
    \raise1pt\hbox{$>$}}}         % greater than or approx. symbol

\def\coordeq{ \, \mathrel{ \rlap{\hbox{\hskip-2.5pt$=$} }
    \raise4pt\hbox{$\cdot$}} \, }                % equal in a particular coordinate system
\begin{document}

\title{A nonlinear scalar model of extreme mass ratio inspirals in effective field theory\\ II. Scalar perturbations and a master source}

\author{Chad R. Galley\footnote{chad.r.galley@jpl.nasa.gov}}

\affiliation{Jet Propulsion Laboratory, California Institute of Technology, Pasadena, CA 91109 USA}
\affiliation{Theoretical Astrophysics, California Institute of Technology, Pasadena, CA 91125 USA}

\date{\today}

\begin{abstract}
	The motion of a small compact (SCO) object in a background spacetime is investigated further in the context of a class of model nonlinear scalar field theories that have a perturbative structure analogous to the General Relativistic description of extreme mass ratio inspirals (EMRIs). We derive regular expressions for the scalar perturbations generated by the motion of the compact object that are valid through third order in $\varepsilon$, the size of the SCO to the background curvature length scale. Our results for the field perturbations are compared to those calculated through second order in $\varepsilon$ by Rosenthal in \cite{Rosenthal:CQG22} and found to agree. However, our procedure for regularizing the scalar perturbations is considerably simpler. Following the Detweiler-Whiting scheme, we use our results for the regular expressions for the field and derive the regular self-force corrections through third order. We find agreement with our previous derivation based on a variational principle of an effective action for the worldline associated with the SCO thereby demonstrating the internal consistency of our formalism. This also explicitly demonstrates that the Detweiler-Whiting decomposition of Green's functions is a valid and practical method of self force computation at higher orders in perturbation theory and, more generally, at all orders in perturbation theory, as we show in an appendix. Finally, we identify a central quantity, which we call a master source, from which all other physically relevant quantities are derivable. Specifically, knowing the master source through some order in $\varepsilon$ allows one to construct the waveform measured by an observer, the regular part of the field and its derivative on the worldline, the regular part of the self force, and various orbital quantities such as shifts of the innermost stable circular orbit, etc., when restricting to conservative dynamics. The existence of a master source together with the regularization methods implemented in this series should be indispensable for derivations of higher-order gravitational self force corrections in the future.
\end{abstract}

\maketitle

%======================================================================
\section{Introduction}
%======================================================================

This paper continues a study begun in \cite{Galley:Nonlinear1} on higher-order self force effects in a nonlinear scalar model as an analog of extreme mass ratio inspirals (EMRIs). In \cite{Galley:Nonlinear1}, henceforth referred to as Paper 1, we derive the finite expressions for the self force on a small compact object (SCO) moving in a background spacetime (such as a supermassive black hole) through third order in the small ratio $\varepsilon \equiv R_m / \cR$. $R_m$ is the typical scale associated with the size of the SCO and $\cR$ is the curvature length scale of the background spacetime. If the SCO with mass $m$ moves in the strong field region of a supermassive black hole with mass $M$ then $\varepsilon \sim m/M$ is the mass ratio of the two bodies.

There are several reasons for studying higher-order self force effects \cite{Galley:Nonlinear1}. These include a need for higher-order self force corrections to maintain the consistency of the inspiral's adiabatic evolution \cite{HindererFlanagan:PRD78} and a need for high-accuracy waveforms to accurately measure the parameters of detected EMRIs in order to perform precise tests of General Relativity with space-based gravitational wave detectors. In addition, the recent discovery of transient resonances, which can significantly dephase the gravitational waveforms if not taken into account, requires modeling EMRI sources with self force corrections through at least second order in $\varepsilon$ \cite{Flanagan:2010cd}. Higher order self force corrections might also be useful to describe binaries with less than extreme mass ratios, such as binaries with intermediate mass ratios in the range of $10^{-1}-10^{-4}$. 

In this paper we continue to focus on higher-order self force effects in a nonlinear scalar model of EMRIs \cite{Galley:Nonlinear1}. Historically, (linear) scalar models offer a simpler framework for studying underlying issues of self force regularization and for developing practical self force computational schemes. Because of the relative simplicities that scalar models afford, it seems likely that the physics of higher-order self force effects can be investigated more easily and more quickly than in the gravitational EMRI context.

In addition to the reasons given above and in Paper 1 for studying higher-order self force effects, our motivation for this second paper comes from several other directions. The first touches on issues of regularity for self force computations. In \cite{DetweilerWhiting:PRD67}, Detweiler and Whiting introduce a physically well-motivated decomposition of the retarded Green's function into regular and singular parts. The regular part satisfies a homogeneous wave equation and is  responsible for generating the regular part of the self force on the SCO. The singular part satisfies an inhomogeneous wave equation, carries all the information about the singular structure (that arises from using a point particle representation for the SCO) and exerts no force on the SCO. 

This decomposition into regular and singular parts is particularly useful for several reasons, including the property that the regular part of the field is differentiable on the worldline. Consequently, one needs only to compute the regular part of the field and its derivative on the worldline in order to evaluate the regular part of the self force. This approach to the self force problem, which we refer to as the ``Detweiler-Whiting (DW) scheme'', is demonstrated in \cite{DetweilerWhiting:PRD67} only at first order in $\varepsilon$. The question naturally arises as to whether or not higher-order self force corrections can be computed in a similar manner. 
We address this question here and explicitly show that the DW scheme for calculating the regular part of the self force holds through third order in $\varepsilon$ and is, in fact, a well-defined and valid procedure at all orders in perturbation theory, at least in the class of nonlinear scalar models considered here.

The second motivation comes from a desire to simplify the regularization and renormalization procedures that tend to impede progress in (higher-order) perturbative calculations of the self force. To our knowledge, Rosenthal is the only one to derive regular expressions for the nonlinear scalar field perturbations generated by the motion of a point particle in a curved, background spacetime through second order in $\varepsilon$ \cite{Rosenthal:CQG22}. As first pointed out using a specific nonlinear field theory in \cite{Rosenthal:CQG22}, scalar perturbations at second order in $\varepsilon$ involve integrals that diverge everywhere in spacetime. Rosenthal's method to obtain well-behaved, finite solutions to the field equation is based on a rather complicated and technical series of steps, which are not so easily generalized to gravitational perturbations because of gauge issues \cite{Rosenthal:PRD73} or to apply at higher-orders in $\varepsilon$. 

The regularization and renormalization methods we use here and in Paper 1 are borrowed from quantum field theory \cite{Peskin}, which has a long history of handling divergences and making sensible, finite predictions that are routinely confirmed by experimental measurements. These techniques are also applicable in curved spacetimes \cite{BirrellDavies}. The theory is well-developed so that there are several practical and effective schemes for regularizing divergent integrals, one of which is dimensional regularization \cite{tHooftVeltman:NuclPhysB44}. Dimensional regularization takes advantage of the observation that the degree of singularity of divergent integrals depends on the number of space-time dimensions, $d$. In some dimensions the integral is actually convergent, which allows one to analytically continue the resulting finite expression for the evaluated integral to the physical number of dimensions\footnote{It is well-known that there may appear singularities when continuing back to $3+1$ dimensions. However, these divergences always manifest as poles in the complex plane of $d$, which can be absorbed into counter terms. In this paper, the analytic continuation of integrals that nominally diverge as a positive power of some cut-off regulator gives zero in $3+1$ dimensions. See \cite{Galley:Nonlinear1, BirrellDavies} for details.}.

We will not use dimensional regularization in most of our calculations in this paper but instead will explicitly carry around the singular integrals to demonstrate that these divergences can be cancelled in a consistent manner by introducing counter terms into the action \cite{Peskin, BirrellDavies}. If, instead, we regulate these singular integrals in dimensional regularization then, as shown in Paper 1, these divergences have a \emph{vanishing} finite part as a result of the analytic continuation implicit in this method. However, even without utilizing dimensional regularization, our approach is considerably simpler and more practical than that of \cite{Rosenthal:CQG22}.

The third motivation is a technical one. The methods of Paper 1 may be unfamiliar to researchers experienced with traditional methods of perturbation theory. Hence, in this paper, we approach the self force problem following the DW scheme. We also wish to show that our formalism is internally consistent and derive the regular part of the self force through third-order in $\varepsilon$ directly from the regular part of the field instead of from a variational principle for the effective action, as in Paper 1. A consistency check of the formalism is that the counter terms derived here are the same as in Paper 1. Indeed, we find precisely the same counter terms and exactly the same expression for the regular self force through third order. 
The DW scheme is a more direct and preferable way to derive self force expressions, in our opinion, than from the effective action of Paper 1. Consequently, we hope the methods used in this paper are more familiar and useful than those in \cite{Galley:Nonlinear1}.

Our last motivation for this second paper comes from what we think is an interesting theoretical observation. We are able to identify what we call a \emph{master source} ${\cal S}$ that appears as the central, fundamental quantity from which most (if not all) other physically relevant quantities are derivable. In particular, knowing the master source through some order in $\varepsilon$ allows one to calculate the (scalar) waveform that an observer would measure in their detector, the regular part of the field anywhere in spacetime (including on the worldline), and the regular part of the self force. In addition, if one restricts oneself to a conservative description for the system then the corresponding master source can be used to compute shifts of the inner-most stable circular orbit, the energy, angular momentum, orbital frequency, and the redshift factor $u^t$ associated with the SCO's orbit, etc. The master source is straightforward to calculate using the methods of this paper. As a demonstration, we derive the expression for the master source through \emph{fourth} order in $\varepsilon$, which one then can use to find regular expressions for the field, the waveform, the self force, etc.

%======================================================================
\subsection{Organization}
%======================================================================

This paper is organized as follows. In Section \ref{sec:nonlinear} we review the nonlinear scalar model of EMRIs, which is meant to serve as an analog of the corresponding gravitational system. In Section \ref{sec:paper1} we highlight the main results from Paper 1 \cite{Galley:Nonlinear1}. In Section \ref{sec:wavesineft} we show how scalar perturbations are computed in the effective field theory (EFT) approach using Feynman diagrams. In Section \ref{sec:waves} we compute the scalar perturbations emitted by the SCO through third order in $\varepsilon$. We also regularize the singular integrals and renormalize the theory by introducing counter terms into the action. In Section \ref{sec:Rosenthal} we compare our results through second order in $\varepsilon$ to those of Rosenthal \cite{Rosenthal:CQG22}. 
In Section \ref{sec:selfforce} we derive the regular part of the self force in the DW scheme, i.e., from the regular part of the scalar field perturbations previously calculated in Section \ref{sec:waves}. We find exactly the same self force expression that we found in Paper 1 thereby providing direct evidence for the consistency of our formalism. In Section \ref{sec:master} we identify the master source and calculate it through \emph{fourth} order in $\varepsilon$. 

We work in units where $G=c=1$, in which case the small parameter for perturbation theory is $\varepsilon \sim m / \cR$ if $m \sim R_m$ is the mass of the small compact object. Throughout, we ignore finite-size effects (such as tidal moments induced on the SCO) and  spin angular momentum for the SCO, which will be considered in future work. Notation is as explained in Paper 1.

%======================================================================
\subsection{The nonlinear scalar model of EMRIs}
\label{sec:nonlinear}
%======================================================================

In Paper I we introduced a class of nonlinear scalar theories designed to have the same structure as the perturbative description of EMRIs in General Relativity. Specifically, if the gravitational waves are in the Lorenz gauge and the SCO moves in a vacuum background spacetime, then the nonlinear scalar model that most resembles the perturbative structure of gravitational EMRIs has an action given by \cite{Galley:Nonlinear1}
\begin{align}
	S [ z^\mu, \phi] = - \frac{ 1}{2} \int_x \phi_{,\alpha} \phi^{,\alpha} A^2 (\phi  ) - m \int d\tau \, B( \phi )
\label{actionphi1}
\end{align}
Here, $\int_x \equiv \int d^4x \, g^{1/2}(x)$, $\int d\tau \equiv \int_{-\infty}^\infty d\tau$, and $A^2$ and $B$ have a series representation in  $\phi$
\begin{align}
	A^2 (\phi ) & = 1 + \sum_{n=1}^\infty \frac{ 2 a_{n+2} }{ (n+2)! } \phi ^n  
\label{paper1Asq1} \\
	B ( \phi  ) & = 1 + \sum_{n=1}^\infty \frac{ b_n }{ n! } \phi^n 
\label{paper1B1}
\end{align}
where $\{a_n, b_n\}$ are freely specifiable coupling constants or parameters.
The equations of motion derived from (\ref{actionphi1}) are
\begin{align}
	\Box \phi & = - \frac{ A' }{ A } \phi_{,\alpha} \phi ^{,\alpha} + m \int d\tau \, \frac{ \delta^4 (x^\mu - z^\mu (\tau)) }{ g^{1/2} } \frac{ B' }{ A^2 } 
\label{paper1eom1} \\
	a^\mu & = - P^{\mu\nu} \nabla_\nu \ln B (\phi  )
\label{paper1eom2}
\end{align}
where a prime on $A$ or $B$ denotes differentiation with respect to $\phi$.

Interestingly, defining a new field variable $\psi$ that satisfies
\begin{align}
	\psi_{,\alpha} = \phi_{,\alpha} A (\phi )
\label{fieldredef1}
\end{align}
leads to a field theory that is linear in the sense that there are no field self-interactions since such terms proportional to $\psi^n$ for $n>2$ are absent from the ensuing action, 
\begin{align}
	S [ z^\mu, \psi ] = - \frac{1}{2} \int _x \psi_{,\alpha} \psi^{,\alpha} - m \int d\tau \, C( \psi )
\label{actionpsi1}
\end{align}
Here,
\begin{align}
	C ( \psi  ) = 1 + \sum_{n=1}^{\infty} \frac{ c_n }{ n! } \psi^n
\label{Cseries1}
\end{align}
and the $c_n$ can be expressed in terms of the $\{a_n, b_n\}$. The equations of motion for the new field $\psi$ and the worldline follow from (\ref{actionpsi1}),
\begin{align}
	\Box \psi & = m \int d\tau \, \frac{ \delta^4 (x^\mu - z^\mu (\tau)) }{ g^{1/2} } C' (\psi)
\label{eom1} \\
	a^\mu & = - P^{\mu\nu} \nabla_\nu \ln C ( \psi ) 
\label{eom2}
\end{align}
where a prime on $C$ denotes differentiation with respect to $\psi$.

%======================================================================
\subsection{Selected results from Paper I}
\label{sec:paper1}
%======================================================================

In Paper 1, we introduced a new action principle, based on (\ref{actionpsi1}), that consistently incorporated the outgoing boundary conditions imposed on solutions to the field equation when integrating out the field to obtain the effective action. The variation of the resulting effective action yielded the worldline equations of motion and hence the self force to the given order in $\varepsilon$.
Specifically, expressions for the self force through third order in $\varepsilon$ were derived in \cite{Galley:Nonlinear1} (in terms of the retarded propagator, which is generally unknown in closed form for most spacetimes) and found to contain singular contributions that depended on the worldline's past history. 
These divergences were shown to vanish in dimensional regularization. However, the formal expressions for those divergences were carried explicitly through the calculations to demonstrate that they could be absorbed into counter terms for the appropriate parameters of the theory. Specifically, these singular pieces were removed by adding the following three local counter terms to (\ref{actionpsi1}),
\begin{align}
	- \delta_m \int d\tau  - \delta_1 \int d\tau \, \psi (z^\mu ) - \frac{ \delta _2 }{ 2 } \int d\tau \, \psi^2 (z^\mu)
\end{align}
where
\begin{align}
	\delta_m = {} & \frac{ m^2 c_1^2 }{ 2  } \bigg( \frac{ \Lambda }{ 4\pi} \bigg) + \frac{ m^3 c_1^2 c_2 }{ 2  } \bigg( \frac{ \Lambda }{ 4\pi} \bigg)^2 + \frac{ m^4 c_1^2 c_2^2 }{ 2  } \bigg( \frac{ \Lambda }{ 4\pi} \bigg)^3 + \frac{ m^4 c_1^3 c_3 }{ 6  } \bigg( \frac{ \Lambda }{ 4\pi} \bigg)^3
\label{deltam1} \\
	\delta_1 = {} &  m^2 c_1 c_2 \bigg( \frac{ \Lambda }{ 4\pi} \bigg) + m^3 c_1 c_2^2 \bigg( \frac{ \Lambda }{ 4\pi} \bigg)^2 + \frac{ m^3 c_1^2 c_3 }{ 2} \bigg( \frac{ \Lambda }{ 4\pi} \bigg)^2 
\label{delta11} \\
	\delta _2 = {} & m^2 c_2^2 \bigg( \frac{ \Lambda }{ 4\pi} \bigg) + m^2 c_1 c_3  \bigg( \frac{ \Lambda }{ 4\pi} \bigg)
\label{delta21}
\end{align}
Then, the remaining regular part of the self force (using the Detweiler-Whiting decomposition for the retarded Green's function \cite{DetweilerWhiting:PRD67}) was found to be
\begin{align}
	F_R^\mu (\tau) = {} & \big( a^\mu + P^{\mu\nu} \nabla_\nu \big) \bigg\{ m^2 c_1^2  I_R (z^\mu) - m^3 c_1^2 c_2 \bigg( \frac{ 1}{2} I_R^2 (z^\mu) + \int d\tau' \, D_R (z^\mu, z^{\mu'}) I_R ( z^{\mu'}) \bigg) \nonumber \\
		& + m^4 c_1^2 c_2^2  \bigg( I_R (z^\mu) \int d\tau' \, D_R (z^\mu, z^{\mu'}) I_R (z^{\mu'}) + \int d\tau' d\tau'' \, D_R (z^\mu, z^{\mu'}) D_R (z^{\mu'}, z^{\mu''}) I_R (z^{\mu''}) \bigg) \nonumber  \\
		& + \frac{ m^4 c_1^3 c_3 }{ 2 } \bigg( \frac{ 1}{3} I_R^3 (z^\mu) + \int d\tau' \, D_R (z^\mu, z^{\mu'}) I_R^2 (z^{\mu'}) \bigg) + O( \varepsilon^4)
\label{paper1sf1}
\end{align}
Setting $F_R^\mu$ equal to $m a^\mu$ yielded the equations of motion for the SCO through third order in $\varepsilon$.

We also showed in Paper 1 that collecting all the terms proportional to the acceleration together gave a compact representation for the worldline equations of motion (i.e., self force)
\begin{align}
	m \Gamma (z^\mu) a^\mu = - m P^{\mu\nu} \nabla_\nu \Gamma (z^\mu)
\label{paper1sf2}
\end{align}
where
\begin{align}
	\Gamma (z^\mu ) = {} & 1 - m c_1^2  I_R (z^\mu) + m^2 c_1^2 c_2  \bigg( \frac{1}{2} I_R^2 (z^\mu) + \int d\tau' \, D_R (z^\mu, z^{\mu'}) I_R (z^{\mu'}) \bigg) \nonumber \\
	& - m^3 c_1^2 c_2^2 \bigg( I_R (z^\mu) \int d\tau' \, D_R (z^\mu, z^{\mu'}) I_R (z^{\mu'}) + \int d\tau' d\tau'' \, D_R (z^\mu, z^{\mu'}) D_R (z^{\mu'}, z^{\mu''}) I_R (z^{\mu''}) \bigg) \nonumber \\
	& - \frac{ m^3 c_1^3 c_3 }{ 2 } \bigg( \frac{ 1}{3} I_R^3 (z^\mu) + \int d\tau' \, D_R (z^\mu, z^{\mu'}) I_R^2 (z^{\mu'}) + O(\varepsilon^4) 
\end{align}
Equivalently, dividing both sides of (\ref{paper1sf2}) by $m \Gamma (z^\mu)$ gave
\begin{align}
	a^\mu = - P^{\mu\nu} \nabla_\nu \ln \Gamma (z^\mu)
\label{paper1sf3}
\end{align}
so that a comparison with (\ref{eom2}) implied $\Gamma (z^\mu) = C ( \psi_R (z^\mu) )$ that, if true, suggested that the regular part of the field would be given by
\begin{align}
	 \psi _R (x) = {} & -  m c_1 I_R (x) + m^2 c_1 c_2 \int d\tau \, D_R (x, z^\mu) I_R (z^\mu) - m^3 c_1 c_2^2 \int d\tau d\tau' \, D_R (x, z^\mu) D_R (z^\mu, z^{\mu'}) I_R (z^{\mu'}) \nonumber \\
	& - \frac{ m^3 c_1^2 c_3 }{ 2 } \int d\tau \, D_R (x, z^\mu) I_R^2 (z^\mu) + O( \varepsilon^4)
\label{paper1Rfield1}
\end{align}
We will show in this paper that $\Gamma(z^\mu) = C(\psi_R (z^\mu))$ and that (\ref{paper1Rfield1}) is indeed the correct expression for the regular part of the field through third order in $\varepsilon$.

%======================================================================
\section{Scalar perturbations in the EFT approach}
\label{sec:wavesineft}
%======================================================================

In this section, we describe how scalar field perturbations are computed within the EFT framework via diagrammatic methods. Perturbations of the scalar field $\psi(x)$ are the perturbative solutions to the wave equation (\ref{eom1}) with a source due to the coupling between the field and the worldline for the SCO
\begin{align}
	\Box \psi & = m \int d\tau \, \frac{ \delta^4 (x^\mu - z^\mu (\tau)) }{ g^{1/2} } \bigg( c_1 + c_2 \psi + \frac{1}{2} c_3 \psi^2 + \cdots \bigg)
\label{waveeqn1}
\end{align}
If $D(x,x')$ is the Green's function, or ``propagator'', appropriate to the problem (e.g., for outgoing boundary conditions) then the solution to (\ref{waveeqn1}) is formally
\begin{align}
	\psi(x) = - m c_1 \int d\tau \, D(x, z^\mu) - m c_2 \int d\tau \, D( x, z^\mu) \psi (z^\mu) - \frac{1}{2} m c_3 \int d\tau \, D(x, z^\mu) \psi^2 (z^\mu) + \cdots
\end{align}
The perturbative (but divergent) solution can be found by solving the above expression iteratively, which through second order in $\varepsilon$ is
\begin{align}
	\psi(x) = - m c_1 \int d\tau \, D(x, z^\mu) + m^2 c_1 c_2 \int d\tau d\tau' D(x, z^\mu) D( z^\mu, z^{\mu'}) + O(\varepsilon^3)
\label{traditional1}
\end{align}
with the higher-order terms becoming increasingly more complicated and more involved to compute\footnote{It is not really the case here that the scalar perturbations are difficult to calculate through third order in $\varepsilon$. However, calculating perturbative solutions for metric perturbations to some order in $\varepsilon$ quickly becomes nontrivial and quite involved, even for relatively low orders. We remind the reader that one purpose of this nonlinear scalar model is to provide useful insights into calculating higher-order self force corrections in the full gravitational EMRI case.}. In addition, one has to carry around many terms just to find the $O(\varepsilon^n)$ contribution for some $n$.
For these reasons, it is beneficial to perturbatively solve the wave equation (\ref{waveeqn1}) using Feynman diagrams, which allow one to systematically compute only those contributions appearing at a given order without the need to carry around many terms that may otherwise be irrelevant at that order. 

Feynman diagrams are constructed by drawing all connected, tree (i.e., not containing any loops of propagator lines, which correspond to quantum corrections) diagrams that scale with $\varepsilon$ to the given power. A diagram representing a field perturbation must have one ``dangling'' propagator (i.e., curly) line having one end that is not attached to a worldline. The free end of that line represents the field point where one is measuring the field. 
To determine the scaling of a diagram, which is a procedure called power counting, we recall that the scalar perturbations vary with the scale of the background spacetime curvature so that $
\partial_\alpha \psi /  \psi \sim \cR^{-1}$ and hence $x^\mu \sim \cR$. In units where $G=c=1$, the action has dimensions of $({\rm Length})^2$ from which it follows that $\psi$ is dimensionless in $3+1$ spacetime dimensions and $\psi \sim \cR^0$. The retarded propagator $D_{\rm ret}(x,x')$ satisfies
\begin{align}
	\int_x \Box D_{\rm ret} (x,x') = - 1 \sim \cR^{0}
\end{align}
from which it follows that $D_{\rm ret} (x,x') \sim \cR^{-2}$. Lastly, every worldline vertex in a diagram scales as $\sim m$ (see first of the Feynman rules listed in Appendix \ref{app:feynman}) and $\int d\tau \sim \cR$ is the typical dynamical time-scale for the motion of the SCO in the background spacetime. These scaling laws allow one to determine which order in $\varepsilon$ that a Feynman diagram appears in perturbation theory. An example applying the Feynman rules and power counting to a specific Feynman diagram is given in Appendix \ref{app:feynman}.

The fact that the perturbative solution for $\psi(x)$ in (\ref{traditional1}) can be expressed via Feynman diagrams is not so surprising. (This is well-known from quantum field theory.) In fact, (\ref{traditional1}) can be written more suggestively as
\begin{align}
	\psi(x) = \int d\tau \, D(x, z^\mu) (- m c_1) + \int d\tau \, D(x, z^\mu) (-m c_2) \int d\tau' \, D(z^\mu, z^{\mu'}) (-m c_1) + O(\varepsilon^3)
\label{traditional2}
\end{align}
which is exactly the expression found by applying\footnote{Start from the top of the propagator line in each figure, which is at the field point $x^\mu$, and work through the diagram, daisy-chaining propagators via worldline vertices and introducing the corresponding factors as stated in the Feynman rules in Appendix \ref{app:feynman}.} the Feynman rules to Figures \ref{fig:firstorderwave} and \ref{fig:secondorderwave} -- see (\ref{field1b}) and (\ref{field2b}) below.

%======================================================================
\section{Scalar perturbations in the nonlinear model}
\label{sec:waves}
%======================================================================

In this section we calculate the regular part of the scalar perturbations emitted by the SCO through third order in $\varepsilon$ within the nonlinear scalar model. We show that all divergences can be removed by introducing suitable counter terms in the action. These counter terms will be shown to precisely equal those of Paper 1 despite the different contexts in which they appear. This thereby demonstrates the internal consistency of our regularization and renormalization program.

%======================================================================
\subsection{Scalar perturbations at first order}
%======================================================================

\begin{figure}
	\includegraphics[width=4cm]{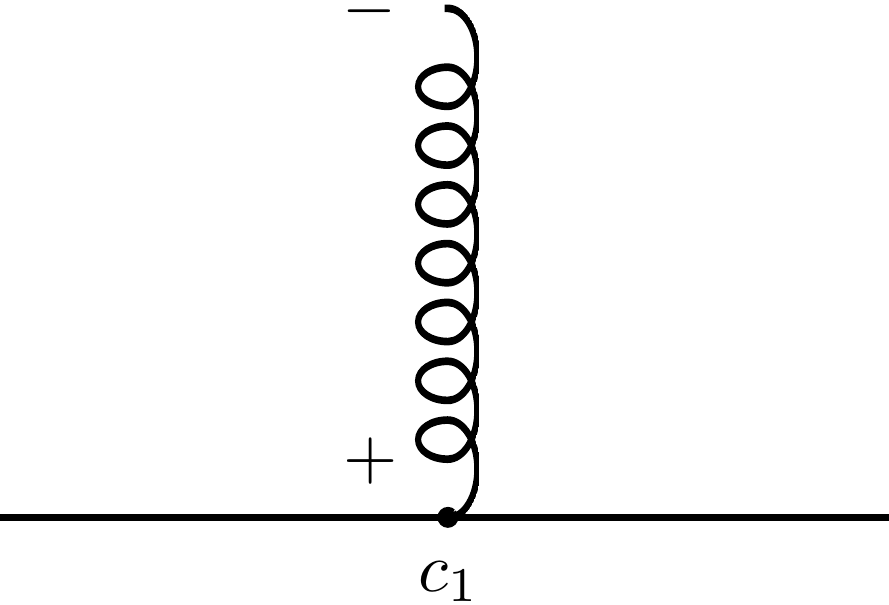}
	\caption{The Feynman diagram contributing to the radiated scalar perturbations at first order in $\varepsilon$. This diagram scales as $(\cR^{-2})(m)(\cR)=\varepsilon$ where the $m$ comes from the vertex, the propagator line scales as $1/\cR^2$, and the proper time integration scales as $\int d\tau \sim \cR$.}
	\label{fig:firstorderwave}
\end{figure}

The leading order contribution to the radiated scalar perturbations comes from the diagram shown in Figure \ref{fig:firstorderwave}. According to the power counting rules this diagram scales linearly with $\varepsilon$ and yields the first-order contribution to the field. 
The Feynman rules imply that Figure \ref{fig:firstorderwave} is equal to
\begin{align}
	\psi_{(1)}(x) & = \int d\tau \, D^{-+}(x, z^\mu) (-m c_1) \\
		& = - m c_1 \int d\tau' \, D_{\rm ret} (x, z^{\mu'}) 
\label{field1b}
\end{align}
where $D^{-+}(x,x') \equiv D_{\rm ret} (x,x')$.
Since we are assuming that the field point $x^\mu$ is not on the worldline then the above integral is manifestly finite and describes the leading order scalar perturbations generated by the motion of the SCO as it moves through the background spacetime.

%======================================================================
\subsection{Scalar perturbations at second order}
%======================================================================

\begin{figure}
	\includegraphics[width=4cm]{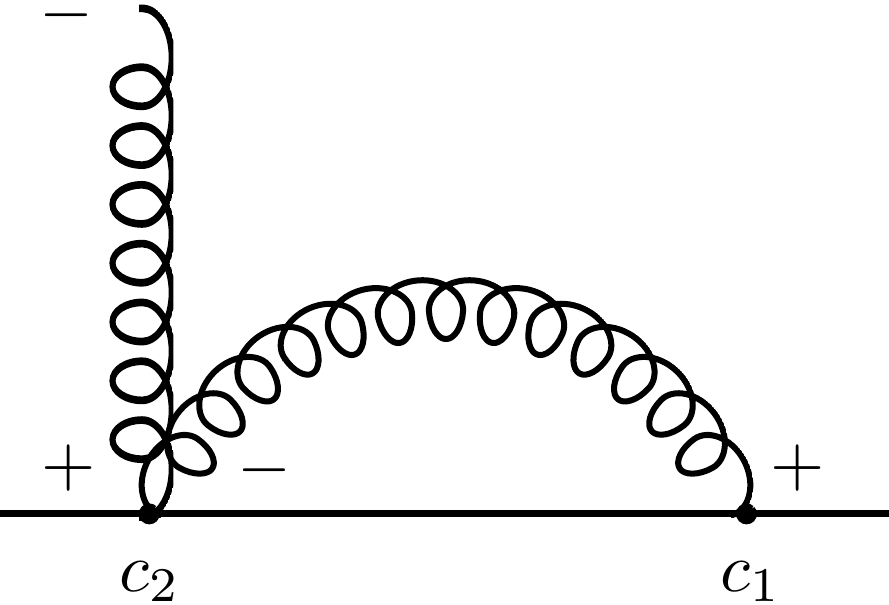}
	\caption{The Feynman diagram contributing to the radiated scalar perturbations at second order in $\varepsilon$. This diagram scales as $(\cR^{-2})(m)(\cR^{-2})(m)(\cR)^2 = \varepsilon^2$.}
	\label{fig:secondorderwave}
\end{figure}

The diagram in Figure \ref{fig:secondorderwave} scales as $\varepsilon^{2}$ and describes the scalar perturbations at second order. The Feynman rules imply that
\begin{align}
	\psi_{(2)} (x) & = \int d\tau' d\tau'' \, D^{-+}(x, z^{\mu'}) (-m c_2)  D^{-+} (z^{\mu'}, z^{\mu''}) (-m c_1) \\
		& = m^2 c_1 c_2 \int d\tau' d\tau'' \, D_{\rm ret} (x, z^{\mu'}) D_{\rm ret} (z^{\mu'}, z^{\mu''})
\label{field2b}
\end{align}
(Compare (\ref{field2b}) with the second order term in (\ref{traditional1}), which is found in a more traditional way.) The integral over $\tau''$ diverges when $\tau'' = \tau'$. 

To deal with this singularity, we use the Detweiler-Whiting prescription \cite{DetweilerWhiting:PRD67} where the retarded Green's function is decomposed into a regular (R) and singular (S) part,
\begin{align}
	D_{\rm ret} (x,x') = D_R (x,x') + D_S (x,x')
\end{align}
While there are many ways to isolate the singular part from the regular one, this particular decomposition has the advantage that the regular part $D_R$ satisfies the homogeneous wave equation and is the piece of the retarded Green's function actually responsible for describing the force on the SCO. However, the singular part $D_S$ carries all the divergent structure of $D_{\rm ret}$, satisfies the inhomogeneous wave equation and exerts no force on the SCO.

In Paper I, we showed how to use the Detweiler-Whiting decomposition to isolate the singular part(s) of the worldline integrals encountered there. We refer the reader to Paper I for the details. To evaluate the divergent integral in (\ref{field2b}) we recall from Paper I that
\begin{align}
	\int d\tau'' \, D_{\rm ret} (z^{\mu'}, z^{\mu''}) = \frac{ \Lambda }{ 4\pi } + I_R (z^{\mu'})
\label{singularintegral1}
\end{align}
where 
\begin{align}
	\frac{\Lambda}{4\pi}  = \int_{-\infty}^\infty d\tau' \, D_S (z^{\mu}, z^{\mu'}) = \int ^{\infty}_0 \!\!\! ds \, \frac{ \delta (s) }{ 4\pi | s | }
\label{lambda1}
\end{align}	
is the singular integral and 
\begin{align}
	I_R (x ^\mu ) \equiv \int_{-\infty}^\infty d\tau'' \, D_R (x^\mu, z^{\mu''})
\label{regularintegral1}
\end{align}
is the regular part of (\ref{singularintegral1}). Then, (\ref{field2b}) equals
\begin{align}
	\psi_{(2)} = m^2 c_1 c_2 \bigg( \frac{ \Lambda }{ 4\pi} \bigg) \int d\tau' \, D_{\rm ret} (x, z^{\mu'} ) + m^2 c_1 c_2 \int d\tau' \, D_{\rm ret} (x, z^{\mu'} ) I_R (z^{\mu'})
\end{align}

%======================================================================
\subsection{Scalar perturbations at third order}
%======================================================================

\begin{figure}
	\includegraphics[width=6.5cm]{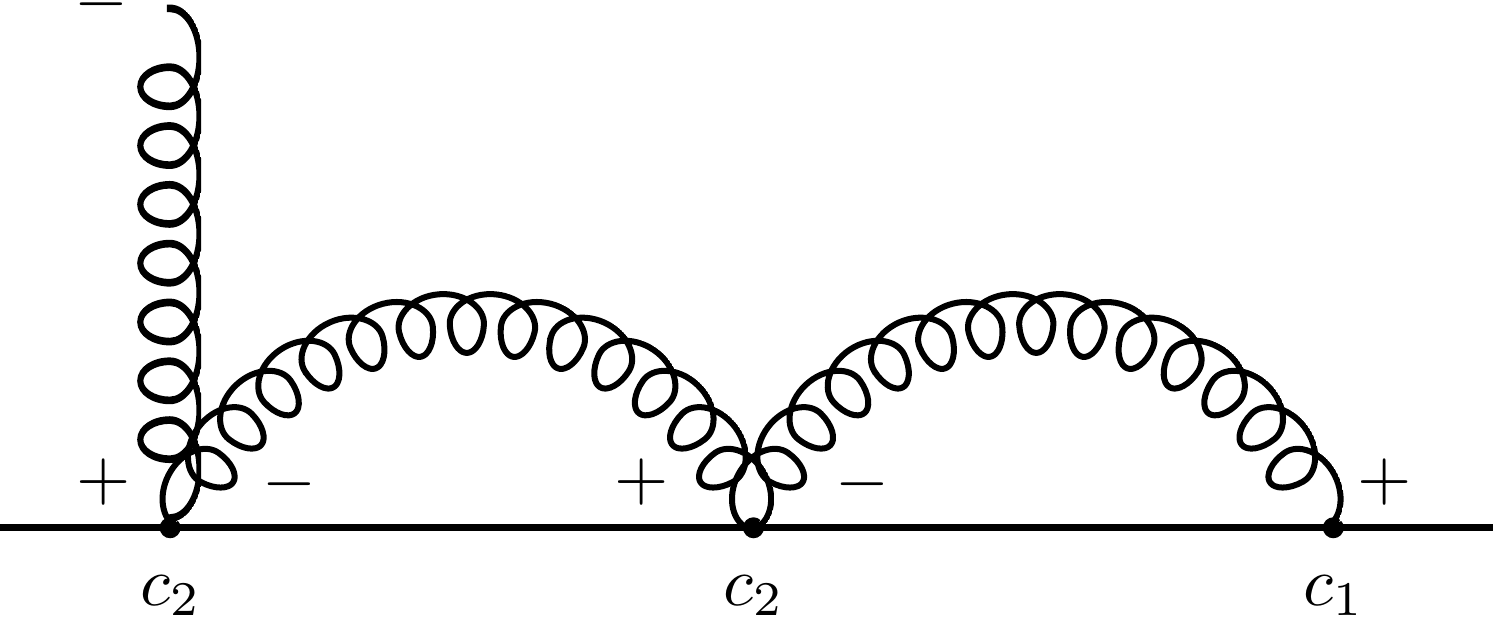} ~~
	\includegraphics[width=6.5cm]{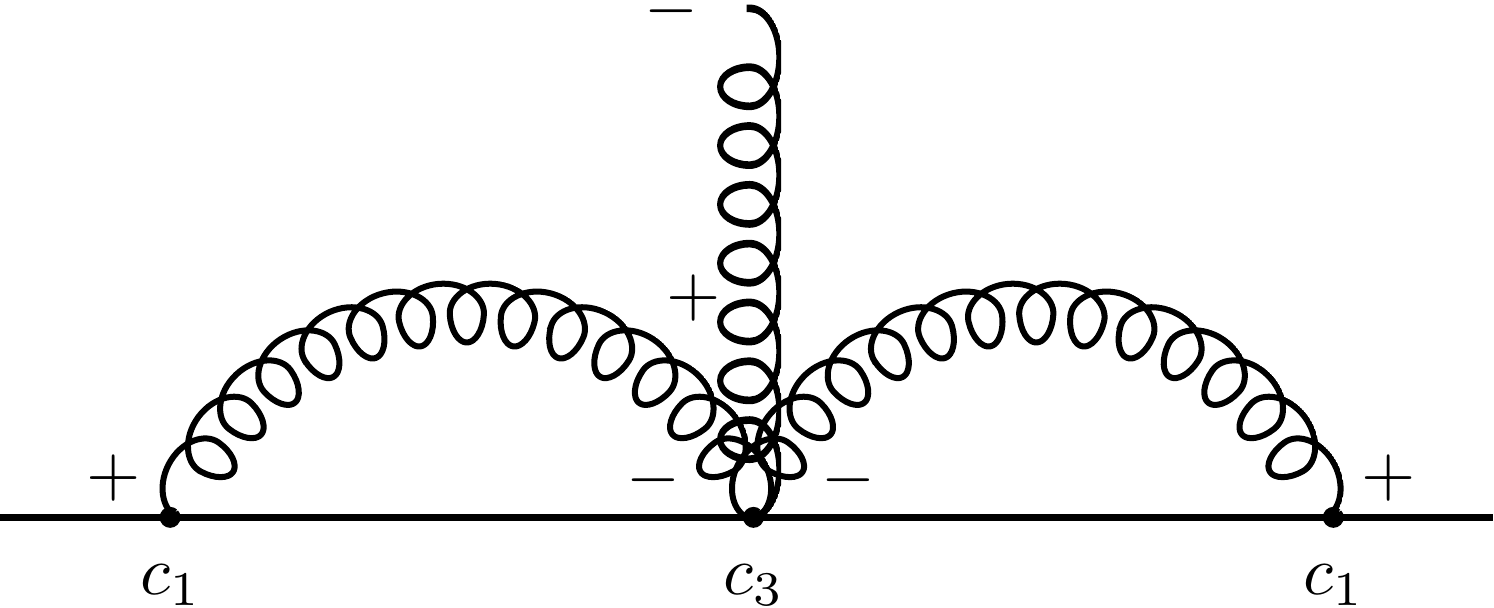}
\caption{The diagrams contributing to the radiated scalar perturbations at third order in $\varepsilon$. Both diagrams scale as $(m)^3 (\cR^{-2})^3 (\cR)^3 = \varepsilon^3$}
\label{fig:thirdorderwave}
\end{figure}

At third order in $\varepsilon$ there are two diagrams contributing to the emitted field, which are shown in Figure \ref{fig:thirdorderwave}. The Feynman rules imply that
\begin{align}
	\psi_{(3)} (x) = {} & \int d\tau' d\tau'' d\tau''' \, D^{-+} (x, z^{\mu'}) (-m c_2)  D^{-+}(z^{\mu'}, z^{\mu''}) (-m c_2) D^{-+}( z^{\mu''}, z^{\mu'''}) (-m c_1) \nonumber \\
		& + \frac{1}{2!} \int d\tau' d\tau'' d\tau''' \, D^{-+ } (x, z^{\mu'}) (-m c_3) D^{-+}(z^{\mu'}, z^{\mu''}) (-m c_1) D^{-+} (z^{\mu'}, z^{\mu'''}) (-m c_1)
\label{field3a}
\end{align}
where the factor of $1/2!$ in the second term comes from the symmetry factor of the diagram on the right in Figure \ref{fig:thirdorderwave}. With $D^{-+}(x,x') = D_{\rm ret} (x,x')$ it follows that
\begin{align}
	\psi_{(3)} (x) = {} & - m^3 c_1 c_2^2 \int d\tau' \, D_{\rm ret} (x, z^{\mu'} ) \int d\tau'' d\tau''' \, D_{\rm ret} (z^{\mu'}, z^{\mu''}) D_{\rm ret} (z^{\mu''} , z^{\mu'''}) \nonumber \\
	& - \frac{ 1 }{ 2 } m^3 c_1^2 c_3 \int d\tau' \, D_{\rm ret} (x, z^{\mu'}) \bigg[ \int d\tau'' \, D_{\rm ret} (z^{\mu'}, z^{\mu''} ) \bigg]^2
\label{field3b}
\end{align}
From Paper I we recall that
\begin{align}
	\int d\tau'' d\tau''' \, D_{\rm ret} (z^{\mu'} , z^{\mu''}) D_{\rm ret} (z^{\mu''}, z^{\mu'''}) = \bigg( \frac{ \Lambda }{ 4 \pi} \bigg)^2 + 2 \bigg( \frac{ \Lambda }{ 4\pi} \bigg) I_R (z^{\mu'}) + \int d\tau'' \, D_R(z^{\mu'} z^{\mu''}) I_R (z^{\mu''})
\label{singularintegral2}	
\end{align}
Therefore, using (\ref{singularintegral2}) in the first line of (\ref{field3b}) and (\ref{singularintegral1}) in the second line gives the following expression for the contribution to the field at third order,
\begin{align}
	\psi_{(3)} = {} & - m^3 c_1 c_2^2 \int d\tau' \, D_{\rm ret} (x,z^{\mu'}) \bigg\{ \bigg( \frac{ \Lambda }{ 4\pi} \bigg)^2 + 2 \bigg( \frac{ \Lambda }{ 4\pi} \bigg) I_R (z^{\mu'}) + \int d\tau'' \, D_R(z^{\mu'}, z^{\mu''}) I_R(z^{\mu''}) \bigg\} \nonumber \\
	& - \frac{ 1}{ 2 } m^3 c_1^2 c_3 \int d\tau' \, D_{\rm ret} (x, z^{\mu'} ) \bigg\{ \bigg( \frac{ \Lambda }{ 4\pi} \bigg)^2 + 2 \bigg( \frac{ \Lambda }{ 4\pi} \bigg) I_R (z^{\mu'}) + I_R^2 (z^{\mu'}) \bigg\}
\label{field3c}
\end{align}

%======================================================================
\subsection{Renormalization}
%======================================================================

Combining our results for the first, second and third order contributions to the scalar perturbations emitted by the SCO from (\ref{field1b}), (\ref{field2b}), and (\ref{field3c}) gives the solution to (\ref{waveeqn1}) through third order in $\varepsilon$,
\begin{align}
	\psi (x) = {} & \int d\tau' \, D_{\rm ret} (x,z^{\mu'}) \bigg\{ m^2 c_1 c_2 \bigg( \frac{ \Lambda }{ 4\pi } \bigg) -  m^3 c_1 c_2^2  \bigg( \frac{ \Lambda }{ 4\pi} \bigg)^2 - \frac{ m^3 c_1^2 c_3 }{ 2 } \bigg( \frac{ \Lambda }{ 4\pi} \bigg)^2 \bigg\} \nonumber \\
	& + \int d\tau' \, D_{\rm ret} (x, z^{\mu'}) I_R (z^{\mu'}) \bigg\{ -  2 m^3 c_1 c_2^2  \bigg( \frac{ \Lambda }{ 4\pi } \bigg) - m^3 c_1^2 c_3 \bigg( \frac{ \Lambda }{ 4\pi } \bigg) \bigg\} \nonumber \\
	& + \int d\tau' \, D_{\rm ret} (x, z^{\mu'}) \bigg\{ - m c_1  + m^2 c_1 c_2 I_R (z^{\mu'}) - m^3 c_1 c_2^2 \int d\tau'' \, D_R (z^{\mu'}, z^{\mu''}) I_R (z^{\mu''}) - \frac{ m^3 c_1^2 c_3 }{ 2 } I_R^2 (z^{\mu'}) \bigg\} \nonumber \\
	&+ O(\varepsilon^4)
\label{field4}
\end{align}

\begin{figure}
	\center
	\includegraphics[width=4cm]{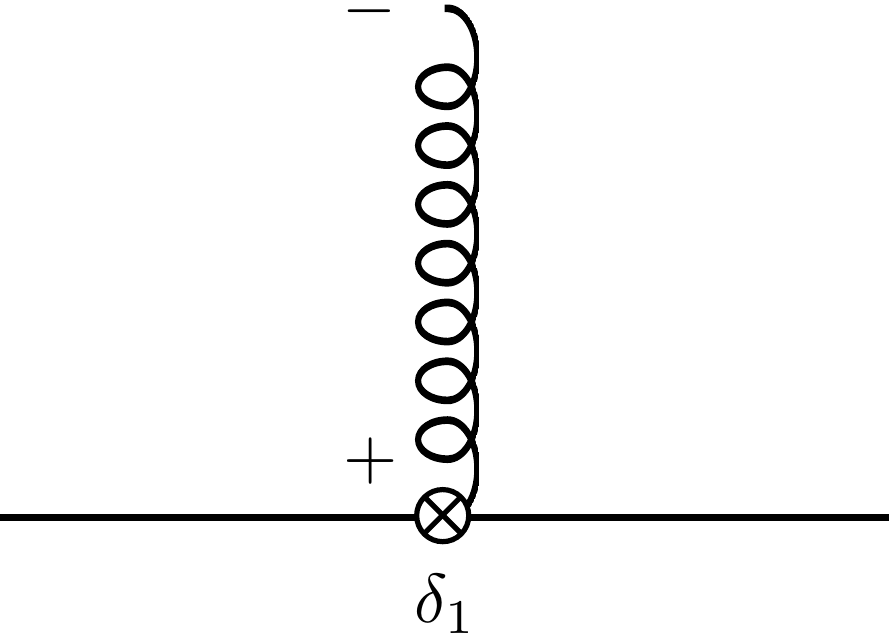}
	\includegraphics[width=4cm]{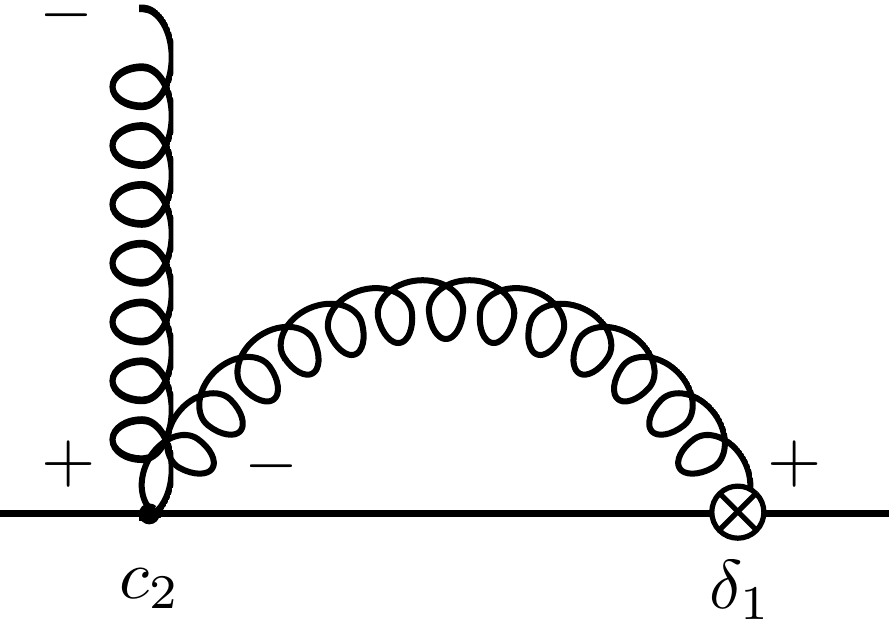}
	\includegraphics[width=4cm]{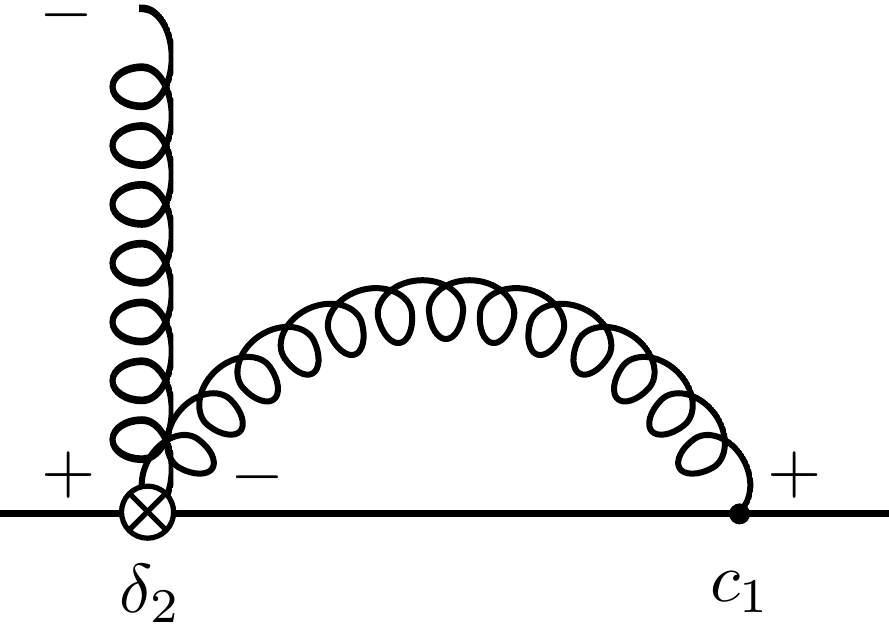} 
\caption{The contributions of the counter terms to the radiated scalar perturbations through third order in $\varepsilon$.}
\label{fig:countertermwaves}
\end{figure}

As with the self force calculation in Paper 1, we find singular contributions to the radiated field that are of higher-order (e.g., $\sim \Lambda^2$) and history-dependent (e.g., $\sim \Lambda I_R (z^{\mu'})$). Despite the appearance of these seemingly pathological structures we can cancel their contributions by adding purely {\it local} counter terms into the action. In particular, we add to the action (\ref{actionpsi1}) the following counter terms
\begin{align}
	- \delta_1 \int d\tau \, \psi( z^\mu) - \frac{ \delta _2 }{ 2 } \int d\tau \, \psi^2 (z^\mu)
	\label{counterterms1}
\end{align}
since these are the ones at lowest orders in $\varepsilon$ that describe a coupling between the worldline and the scalar perturbation. Consequently, the counter terms modify the equations of motion and their (divergent) solutions so as to make them finite. The contributions to the radiated scalar perturbations from these counter terms are displayed in Figure \ref{fig:countertermwaves}.

Following steps similar to those used in deriving (\ref{field4}) we find that the contribution of these counter terms to the field is
\begin{align}
	\psi _{\rm CT} (x) = {} & \int d\tau' \, D_{\rm ret} (x, z^{\mu'}) \bigg\{ - \delta_1 + m c_1 \delta_2  \bigg( \frac{ \Lambda }{ 4\pi } \bigg)  + m c_2 \delta_1 \bigg( \frac{ \Lambda }{ 4\pi } \bigg) \bigg\} \nonumber \\
	&  + \int d\tau' \, D_{\rm ret} (x, z^{\mu'}) I_R (z^{\mu'}) \bigg\{ m c_1 \delta_2 + m c_2 \delta_1 \bigg\} + O(\varepsilon^4)
\end{align}
Requiring that $\psi + \psi_{CT}$ be regular and independent of the divergent quantities $\Lambda$ implies that $\delta_1$ and $\delta_2$ are given by
\begin{align}
	\delta_1 = {} & m^2 c_1 c_2 \bigg( \frac{ \Lambda }{ 4\pi } \bigg) + m^3 c_1 c_2^2 \bigg( \frac{ \Lambda }{ 4\pi } \bigg)^2 + \frac{ m^3 c_1^2 c_3}{2 } \bigg( \frac{ \Lambda }{ 4\pi } \bigg)^2 
\label{delta12} \\
	\delta_2 = {} & m^2 c_2^2  \bigg( \frac{ \Lambda }{ 4\pi } \bigg) + m^2 c_1 c_3 \bigg( \frac{ \Lambda }{ 4\pi } \bigg)
\label{delta22}
\end{align}
Therefore, the regular scalar perturbations emitted by the SCO through third order in $\varepsilon$ are
\begin{align}
	\psi_{\rm rad} (x) \equiv {} & \psi(x) + \psi_{\rm CT} (x) \\
		= {} &  \int d\tau' \, D_{\rm ret} (x, z^{\mu'}) \bigg\{ - m c_1 +  m^2 c_1 c_2  I_R (z^{\mu'}) - m^3 c_1 c_2^2  \int d\tau'' \, D_R (z^{\mu'}, z^{\mu''}) I_R (z^{\mu''}) - \frac{ m^3 c_1^2 c_3 }{ 2 } I_R^2 (z^{\mu'}) \bigg\} \nonumber \\
	& + O(\varepsilon^4)
\label{field5}
\end{align}
This is one of our main results. The field $\psi_{\rm rad}$ in (\ref{field5}), which we will call the radiative field for definiteness, is the regular field measured by an observer away from the SCO and, by analogy, corresponds to the gravitational waveform of an EMRI source. It is a strong check of self-consistency that the counter term coefficients $\delta_1$ and $\delta_2$  in (\ref{delta12}) and (\ref{delta22}) have the same values as first derived in Paper I (and given in (\ref{delta11}) and (\ref{delta21})) because of the different ways that they are obtained. Specifically, the counter terms here are introduced to make the field regular while those in Paper 1 to make the self force regular.

For later convenience, we will write (\ref{field5}) in a more compact form,
\begin{align}
	\psi_{\rm rad}(x) = \int d\tau' \, D_{\rm ret} (x, z^{\mu'}) {\cal S}_R (z^{\mu'})
\label{field6}
\end{align}
where the regular source term ${\cal S}_R (z^{\mu'})$ is 
\begin{align}
	{\cal S}_R (z^\mu) \equiv - m c_1 + m^2 c_1 c_2 I_R (z^\mu) - m^3 c_1 c_2^2 \int d\tau' \, D_R (z^\mu, z^{\mu'}) I_R (z^{\mu'}) - \frac{ m^3 c_1^2 c_3 }{2} I_R^2 (z^\mu) + O(\varepsilon^4) 
\label{source1}
\end{align}

%======================================================================
\section{Comparison with Rosenthal's second order scalar perturbations}
\label{sec:Rosenthal}
%======================================================================

Given that (\ref{field5}) is the expression for the regular part of the radiative field and is valid through $O(\varepsilon^3)$, we next compare our result with the $O(\varepsilon^2)$ regular field derived by Rosenthal in \cite{Rosenthal:CQG22}. There, a particular nonlinear scalar field theory is chosen such that it yields the following field equation
\begin{align}
	\Box \phi = \phi_{,\alpha} \phi^{,\alpha} - q \int d\tau \, \frac{ \delta^4 (x^\mu - z^\mu(\tau)) }{ g^{1/2} }
\label{Eran1}
\end{align}
where $q$ denotes the scalar charge of the particle. The challenge in solving this equation, as noted by Rosenthal \cite{Rosenthal:CQG22}, can be seen by decomposing the field as
\begin{align}
	\phi (x) = q \phi_1 (x) + q^2 \phi_2 (x) + O(q^3)
\end{align}
where the subscript labels the order of the perturbative solution. Substituting this into (\ref{Eran1}) and using the solution for the first order perturbation
\begin{align}
	\phi_1 (x)  = \int d\tau \, D_{\rm ret} (x, z^\mu)
\end{align}
gives the following equation for $\phi_2$
\begin{align}
	\Box \phi_2 =  \int d\tau d\tau' \, \nabla_\alpha D_{\rm ret} (x, z^\mu ) \nabla^\alpha D_{\rm ret} (x, z^{\mu'})
\end{align}
The corresponding solution is formally given by 
\begin{align}
	\phi_2 (x) = - \int_y \, D_{\rm ret} (x, y) \int d\tau d\tau'  \, \nabla_\alpha D_{\rm ret} (y, z^\mu) \nabla^\alpha D_{\rm ret} (y, z^{\mu'}) 
\end{align}
which diverges as the inverse second power of the proper spatial distance orthogonal to the worldline as one integrates over $y^\mu$ near the particle. In fact, this divergence arises at every field point $x^\mu$ in the spacetime.

In \cite{Rosenthal:CQG22} Rosenthal devised a series of steps to obtain well-defined solutions to (\ref{Eran1}). Through second order, his prescription yields
\begin{align}
	\phi ( x) = q \int d\tau \, D_{\rm ret} (x, z^\mu) + \frac{ q^2 }{ 2} \bigg[ \int d\tau \, D_{\rm ret} (x, z^\mu) \bigg]^2 - q^2 \int d\tau \, D_{\rm ret} (x, z^\mu) I_{R} (z^\mu) + O(\varepsilon^3)
\label{Eran2}
\end{align}
where $I_R (x)$ is given in (\ref{regularintegral1}).
Rosenthal derived this solution by carefully following several steps: (1) investigating the behavior of the wave equation when $q \to 0$, (2) making an ansatz for the particular solution to the wave equation, (3) using physical considerations to identify the divergent boundary conditions for the field $\phi(x)$ as the field point $x^\mu$ approaches the worldline, and (4) solving the wave equation with the divergent boundary conditions from the previous step. This is a rather tedious process in practice and becomes more involved when calculating second order gravitational perturbations \cite{Rosenthal:PRD73} or going to higher-orders in $\varepsilon$. We show below that one need not employ such a series of steps. Instead, we will recover Rosenthal's perturbative solution for the field (\ref{Eran2}) using effective field theory techniques.

Rosenthal's nonlinear field theory is a member of our class of nonlinear scalar models. To make a meaningful comparison, we need first to find the values of the parameters $\{a_n, b_n\}$ appearing in (\ref{paper1eom1}). From the wave equation in (\ref{paper1eom1}) and the definitions of $A^2(\phi)$ and $B(\phi)$ in (\ref{paper1Asq1}) and (\ref{paper1B1}), respectively, it is easy to show that
\begin{align}
	\frac{ A'}{ A } & = \frac{a_3 }{ 6} + \cdots
\label{AprimeoverA1} \\
	\frac{ B' }{ A^2 } & = b_1 + \bigg( b_2 - \frac{ a_3 b_1}{ 3} \bigg) \phi + \cdots
\label{BprimeoverAsq1}
\end{align}
Inserting (\ref{AprimeoverA1}) and (\ref{BprimeoverAsq1}) into the wave equation for $\phi(x)$ in (\ref{paper1eom1}) and comparing the resulting expression order-by-order in powers of $\phi$ with Rosenthal's wave equation (\ref{Eran1}) implies that
\begin{align}
	b_1 & = - \frac{ q }{ m } \\
	b_2 & = + 2 \frac{ q  }{ m } \\
	a_3 & = -6
\end{align}
and $a_{n\ge 4} = 0$, $b_{n \ge 3} =0$.
Substituting these values for $\{ a_n, b_n \}$ into the renormalized expression for the radiative field given in (\ref{field5}) gives
\begin{align}
	 \psi _{\rm rad} (x) = {} & \int d\tau \, D_{\rm ret} (x, z^\mu) \bigg\{ q - q^2 I_R (z^\mu) + O (\varepsilon^3) \bigg\}
\label{comparepsi1}
\end{align}
Using the expression for $\phi$ in terms of $\psi=\psi_{\rm rad}$ by applying the inverse of the field redefinition of (\ref{fieldredef1}) yields
\begin{align}
	\phi (x) = \psi_{\rm rad} (x) + \frac{1}{2} \psi_{\rm rad}(x)^2 + \cdots
\end{align}
so that substituting in for $\psi_{\rm rad} (x)$ from (\ref{comparepsi1}) gives
\begin{align}
	\phi(x) = q \int d\tau \, D_{\rm ret} ( x, z^\mu) + \frac{ q^2 }{ 2} \bigg[ \int d\tau \, D_{\rm ret} (x, z^\mu) \bigg]^2 - q^2 \int d\tau \, D_{\rm ret} (x, z^\mu) I_R (z^\mu) + O(\varepsilon^3)
\label{Eran3}
\end{align}
Comparing with (\ref{Eran2}) we see that we have reproduced exactly Rosenthal's result \cite{Rosenthal:CQG22}. However, we have arrived at (\ref{Eran3}) using a simpler and more straightforward regularization procedure than that developed in \cite{Rosenthal:CQG22}. 

Our approach to regularization can be streamlined even further by evaluating the singular integrals in dimensional regularization. Since all such integrals are not logarithmically divergent then the divergences will vanish in dimensional regularization. Regularizing in this way then becomes rather trivial.

%======================================================================
\section{Self force from backreaction of scalar perturbations}
\label{sec:selfforce}
%======================================================================

In the previous section, we found the formally divergent solution to the wave equation for $\psi(x)$, given by (\ref{field4}), through third order in $\varepsilon$. The divergences were removed by introducing counter terms into the action, which yielded the regular scalar perturbations radiated to an observer. In this section, we show how to obtain the regular part of the self force through third order from knowledge of $\psi_{\rm rad}(x)$ (i.e., we implement the DW scheme through third order). We also compare with the result we obtained earlier in Paper 1 for the self force. We show in Appendix \ref{app:allorders} that the DW scheme can be applied at any order in perturbation theory.

%======================================================================
\subsection{Regular part of the self force}
%======================================================================

In renormalizing the field $\psi(x)$ in Section \ref{sec:waves} we added the counter terms in (\ref{counterterms1}) to the action (\ref{actionpsi1}),
\begin{align}
	S [ z^\mu, \psi ] + S_{\rm CT} [z^\mu, \psi ] = - \frac{1}{2} \int_x \psi_{,\alpha} \psi^{,\alpha} - \int d\tau \bigg\{ m + (mc_1 + \delta_1) \psi + \frac{1}{2} (m c_2 + \delta _2) \psi^2 + \frac{1}{6} m c_3 \psi^3 + O(\varepsilon^4) \bigg\}
\label{ctaction1}
\end{align}
In doing so the field and worldline equations of motion acquire additional contributions arising from the presence of these counter terms. Specifically, the wave equation becomes
\begin{align}
	\Box \psi (x) =  \int d\tau \, \frac{ \delta ^4 (x^\mu - z^\mu(\tau)) }{ g^{1/2}} \bigg\{ ( m c_1 + \delta_1) + ( m c_2 + \delta_2 ) \psi + \frac{mc_3}{2}  \psi^2  + O( \varepsilon^4) \bigg\}
\label{psiradwaveeqn1}
\end{align}
The resulting solution to this wave equation is precisely $\psi(x) = \psi_{\rm rad}(x)$, the radiative field already given in (\ref{field5}), which can be verified by direct substitution and using (\ref{delta12}) and (\ref{delta22}). However, this should be obvious since the counter terms are introduced in the first place to cancel the divergences in (\ref{field4}) and yield the finite expression in (\ref{field5}), namely, $\psi_{\rm rad}(x)$.

The self force also inherits extra contributions from the counter terms. In addition, the self force that results by varying the action in (\ref{actionpsi1}) with respect to the worldline coordinates is in terms of the radiative field, as this is the quantity that perturbatively solves the wave equation with the counter terms included, (\ref{psiradwaveeqn1}). The resulting self force expression is
\begin{align}
	F^\mu(\tau) = {} & - ( a^\mu + P^{\mu\nu} \nabla_\nu ) \bigg\{ (m c_1 +  \delta_1) \psi_{\rm rad} (z^\mu) + \frac{ 1 }{2} (m c_2 + \delta_2) \psi^2_{\rm rad} (z^\mu)  + \frac{mc_3}{6}  \psi_{\rm rad}^3 (z^\mu) \bigg\} + O(\varepsilon^4)
\label{singularsf1}
\end{align}
upon expanding out $C (\psi)$ using (\ref{Cseries1}). Substituting in (\ref{field5}) for $\psi_{\rm rad}(z^\mu)$ into $F^\mu(\tau)$ in (\ref{singularsf1}) then gives
\begin{align}
	F^\mu(\tau) = {} & F^\mu_R (\tau) + a^\mu \bigg\{ \frac{ m^2 c_1^2 }{ 2 } \bigg( \frac{ \Lambda }{ 4\pi}  \bigg) + \frac{ m^3c_1^2 c_2 }{ 2 } \bigg( \frac{ \Lambda }{ 4\pi}  \bigg)^2 + \frac{ m^4 c_1^2 c_2^2 }{ 2 } \bigg( \frac{ \Lambda }{ 4\pi}  \bigg)^3 + \frac{ m^4 c_1^3 c_3 }{ 6 } \bigg( \frac{ \Lambda }{ 4\pi}  \bigg)^3 + O(\varepsilon^4) \bigg\}
\label{singularsf2}
\end{align}
where $F^\mu_R (\tau)$ is the regular part of the self force through third order,
\begin{align}
	F_R^\mu (\tau) = {} &  \big( a^\mu + P^{\mu\nu} \nabla_\nu \big) \bigg\{ m^2 c_1^2 I_R( z^\mu)  - m^3 c_1^2 c_2 \bigg( \frac{1}{2} I_R^2 (z^\mu)  + \int d\tau' \, D_R (z^\mu, z^{\mu'}) I_R (z^{\mu'}) \bigg) \nonumber \\
	& + m^4 c_1^2 c_2^2 \bigg( I_R (z^\mu) \int d\tau' \, D_R (z^{\mu}, z^{\mu'}) I_R ( z^{\mu'}) + \int d\tau' d\tau''  \, D_R (z^\mu, z^{\mu'}) D_R (z^{\mu'}, z^{\mu''}) I_R (z^{\mu''}) \bigg) \nonumber \\
	& + \frac{ m^4 c_1^3 c_3}{ 2 } \bigg( \frac{ 1}{3}  I_R^3 (z^\mu) + \int d\tau' \, D_R (z^{\mu}, z^{\mu'} ) I_R^2 (z^{\mu'}) \bigg) +O(\varepsilon^4) \bigg\}  
\label{renormalizedsf1}
\end{align}

There are remaining divergent contributions in (\ref{singularsf2}) that are not cancelled by the $\delta_{1}$ and $\delta_2$ counter terms. However, these divergences are all proportional to the particle's acceleration implying that a mass counterterm $\delta_m$ is sufficient to cancel these pieces. Therefore, adding $- \delta_m \int d\tau$ to the action in (\ref{ctaction1}) yields
\begin{align}
	F^\mu (\tau)  - \delta_m a^\mu =  F^\mu _R (\tau) + a^\mu \bigg\{ -\delta_m + \frac{ m^2 c_1^2 }{ 2 } \bigg( \frac{ \Lambda }{ 4\pi}  \bigg) + \frac{ m^3c_1^2 c_2 }{ 2 } \bigg( \frac{ \Lambda }{ 4\pi}  \bigg)^2 + \frac{ m^4 c_1^2 c_2^2 }{ 2 } \bigg( \frac{ \Lambda }{ 4\pi}  \bigg)^3 + \frac{ m^4 c_1^3 c_3 }{ 6 } \bigg( \frac{ \Lambda }{ 4\pi}  \bigg)^3 + O(\varepsilon^4) \bigg\}
\end{align}
so that choosing
\begin{align}
	\delta _m = \frac{ m^2 c_1^2 }{ 2 } \bigg( \frac{ \Lambda }{ 4\pi}  \bigg) + \frac{ m^3c_1^2 c_2 }{ 2 } \bigg( \frac{ \Lambda }{ 4\pi}  \bigg)^2 + \frac{ m^4 c_1^2 c_2^2 }{ 2 } \bigg( \frac{ \Lambda }{ 4\pi}  \bigg)^3 + \frac{ m^4 c_1^3 c_3 }{ 6 } \bigg( \frac{ \Lambda }{ 4\pi}  \bigg)^3 + O(\varepsilon^4) 
	\label{deltam10}
\end{align}
renders the self force finite
\begin{align}
	F^\mu (\tau) - \delta_m a^\mu = F^\mu _R (\tau)
\end{align}
where $F^\mu_R(\tau)$ is given in (\ref{renormalizedsf1}) and is precisely the same expression derived in Paper 1 (shown in this paper in (\ref{paper1sf1})). This is our second main result. In addition, the expression for the mass counter term in (\ref{deltam10}) is exactly the same as given in Paper 1. The agreement among the expressions for all three counter terms and for the regular part of the self force as derived here (via the scalar field perturbations) and in Paper 1 (via a variational principle) provides a strong check of the consistency, results, and tools used in the effective field theory approach for deriving self force corrections in a nonlinear field theory coupled to a worldline.

%======================================================================
\subsection{An efficient derivation of the third order regular self force}
\label{sec:efficient}
%======================================================================

The regular part of the self force in (\ref{renormalizedsf1}) can be derived in a more straightforward way. 
All singular contributions vanish in dimensional regularization (i.e., $\Lambda = 0$), as already noted multiple times here and shown in Paper 1. Therefore, the counter terms $\delta_1$, $\delta_2$, and $\delta_m$ in (\ref{delta12}), (\ref{delta22}), and (\ref{deltam10}) all vanish, the regular part of the field evaluated on the worldline is just 
\begin{align}
	\psi(z^\mu) = \psi_{\rm rad}(z^\mu) = \psi_R (z^\mu) \equiv \int d\tau' \, D_R (z^\mu, z^{\mu'}) {\cal S}_R (z^{\mu'}) ,
\label{regularfield2}
\end{align}
and the formally divergent expression in (\ref{singularsf1}) for the self force becomes manifestly finite. Therefore,
\begin{align}
	ma^\mu = F^\mu (\tau) = F^\mu _R (\tau) = {} & - m \big( a^\mu + P^{\mu\nu} \nabla_\nu \big) \bigg( c_1 \psi_R + \frac{ c_2}{2} \psi^2_R + \frac{ c_3 }{ 6 } \psi_R^3 + O(\varepsilon^4) \bigg)
\label{renormalizedsf2}
\end{align}
where $\psi_R$ is given in (\ref{regularfield2}) and (\ref{source1}), which is subsequently substituted into (\ref{renormalizedsf2}) to recover the same expression for the regular self force through third order given in (\ref{renormalizedsf1}).

%======================================================================
\section{The master source}
\label{sec:master}
%======================================================================

The field $\psi(x)$ in (\ref{field4}) can be written as if there is a source ${\cal S} (z^\mu)$ generating the scalar perturbations where
\begin{align}
	{\cal S} (z^\mu) \equiv {} & m^2 c_1 c_2 \bigg( \frac{ \Lambda }{ 4\pi } \bigg) - m^3 c_1 c_2^2 \bigg( \frac{ \Lambda }{ 4\pi } \bigg)^2 - \frac{ m^3 c_1^2 c_3 }{ 2 } \bigg( \frac{ \Lambda }{ 4\pi } \bigg)^2 - I_R(z^\mu) \bigg\{ - 2 m^3 c_1 c_2^2  \bigg( \frac{ \Lambda }{ 4\pi } \bigg) - m^3 c_1^2 c_3 \bigg( \frac{ \Lambda }{ 4\pi } \bigg) \bigg\} \nonumber \\
	& - m c_1 + m^2 c_1 c_2 I_R (z^\mu) - m^3 c_1 c_2^2 \int d\tau' \, D_R (z^\mu, z^{\mu'} ) I_R (z^{\mu'}) - \frac{ m^3 c_1^2 c_3}{ 2 } I_R^2 (z^\mu) + O(\varepsilon^4)
\end{align}
so that 
\begin{align}
	\psi(x) = \int d\tau' \, D_{\rm ret} (x,z^{\mu'}) {\cal S} (z^{\mu'})
\end{align}
This representation for the field holds to all orders in $\varepsilon$, which is easily seen by solving formally the exact field equation in (\ref{eom1}),
\begin{align}
	\psi(x) = - m \int d\tau \, D_{\rm ret} (x, z^{\mu}) C'(\psi(z^\mu))
\end{align}
implying that
\begin{align}
	{\cal S}(z^\mu) = - m C' (\psi (z^\mu))
\end{align}
For reasons that will become clear below we call ${\cal S}(z^\mu)$ the \emph{master source} function\footnote{We call ${\cal S}(x)$ a ``master source'' since it is representative of all the field-worldline interactions that source the field measurable by some observer. Normally, ${\cal S}$ might be called an ``effective source'' but this term is already used to describe the finite, windowed source term for computing regular perturbations in some approaches to numerical self force computations (e.g., \cite{Vegaetal:PRD80}). ${\cal S}$ might also be called a ``dressed worldline vertex'' as in \cite{KolSmolkin:PRD80}.}.
It will be convenient to identify the regular and singular parts of the master source, ${\cal S} (z^\mu) = {\cal S}_R (z^\mu) + {\cal S}_S (z^\mu)$, with
\begin{align}
	{\cal S}_R (z^\mu) & = - m C' ( \psi_R (z^\mu)) 
\label{mastersrc1} \\
	{\cal S}_S (z^\mu) & = {\cal S} (z^\mu) - {\cal S}_R (z^\mu) \\
\end{align}
Through third order in $\varepsilon$, the regular part of the master source ${\cal S}_R (z^\mu)$ is given in (\ref{source1}) and the singular part by
\begin{align}
	{\cal S}_S (z^\mu) & = m^2 c_1 c_2 \bigg( \frac{ \Lambda }{ 4\pi } \bigg) - m^3 c_1 c_2^2 \bigg( \frac{ \Lambda }{ 4\pi } \bigg)^2 - \frac{ m^3 c_1^2 c_3 }{ 2 } \bigg( \frac{ \Lambda }{ 4\pi } \bigg)^2 - I_R(z^\mu) \bigg\{ -  2 m^3 c_1 c_2^2 \bigg( \frac{ \Lambda }{ 4\pi } \bigg) - m^3 c_1^2 c_3 \bigg( \frac{ \Lambda }{ 4\pi } \bigg) \bigg\}  + O( \varepsilon^4)
\end{align}
In dimensional regularization $\Lambda = 0$ and the singular part of the master source vanishes, ${\cal S}_S = 0$. We will ignore ${\cal S}_S$ in the remainder.

The key point to note is that the master source is the fundamental quantity to calculate for EMRIs: it is used to construct the radiative scalar perturbations observed by detectors (\ref{field6}),
\begin{align}
	\psi_{\rm rad}(x) & = \int d\tau \, D_{\rm ret} (x, z^\mu) {\cal S}_R (z^\mu) ,
\label{scalarrad1}
\end{align}
the regular part of the field (\ref{regularfield2}),
\begin{align}
	\psi_R (x) & = \int d\tau \, D_R (x, z^{\mu}) {\cal S}_R (z^{\mu}) ,
\label{scalarR1}
\end{align}
and, subsequently, the regular self force (\ref{eom2})
\begin{align}
	F^\mu_R (\tau) & = - m (a^\mu + P^{\mu\nu} \nabla_\nu) \sum_{n=1}^\infty \frac{c_n}{n!} \psi^n_R (z^\mu) 
\label{sfR1} 	
\end{align}
where the self force depends on ${\cal S}_R(z^\mu)$ through (\ref{scalarR1}) evaluated on the worldline. Also, the physical field is $\phi_{\rm rad}(x)$ and the corresponding regular part on the worldline is $\phi_R (z^\mu)$, both of which can be calculated by inverting the field redefinition in (\ref{fieldredef1}) in terms of $\psi_{\rm rad}(x)$ and $\psi_R(z^\mu)$, respectively, which both depend on the master source. 
Therefore, by computing ${\cal S}_R (z^\mu)$ to the desired order in $\varepsilon$ one can compute essentially all relevant physical quantities describing the EMRI. This is our third main result. In the next subsection we show \emph{how} to calculate the regular master source using diagrammatic techniques.

%======================================================================
\subsection{The master source from Feynman diagrams through third order}
%======================================================================

\begin{figure}
\subfigure[]{
	\includegraphics[width=4cm]{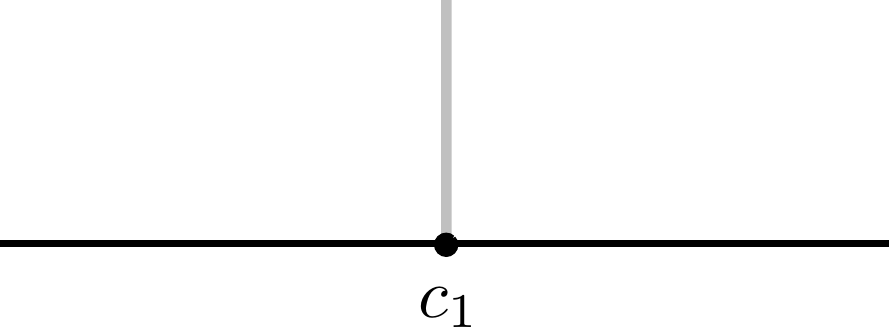} 
	\label{fig:VR1}
}
\subfigure[]{
	\includegraphics[width=4cm]{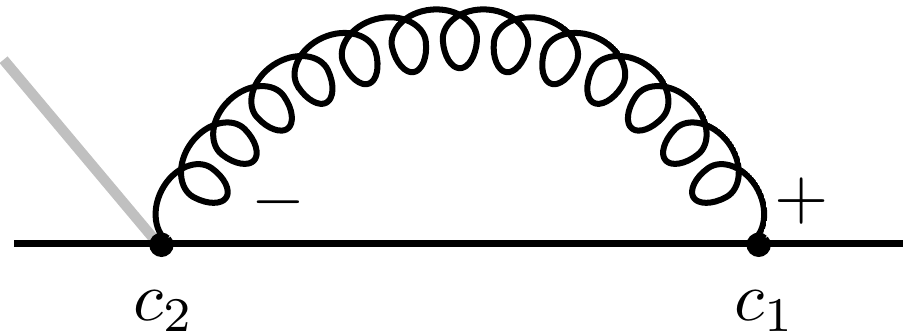} 
	\label{fig:VR2}
}
\subfigure[]{
	\includegraphics[width=6.5cm]{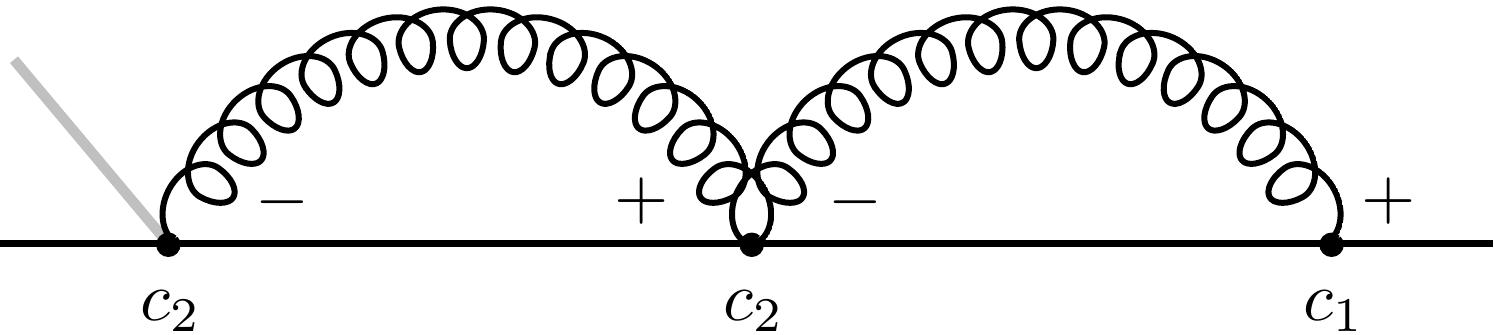} ~~
	\includegraphics[width=6.5cm]{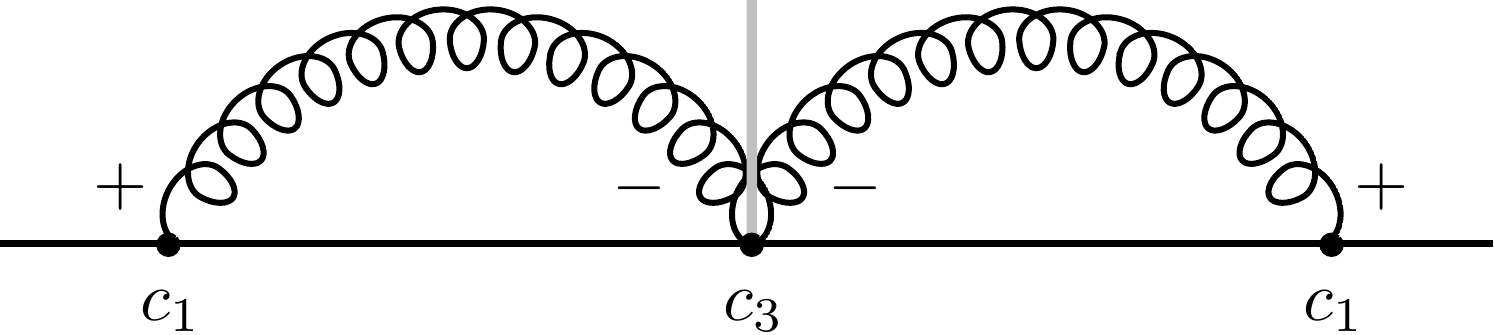} 
	\label{fig:VR3}
}
\caption{The diagrams contributing to the regular part of the master source ${\cal S}_R( z^\mu)$ at (a) first order ($\sim \cR \varepsilon$), (b) second order ($\sim \cR \varepsilon^2$), and (c) third order ($\sim \cR \varepsilon^3$) in $\varepsilon$. The solid gray line emanating from a vertex represents the place to attach a Green's function such as $D_{\rm ret}$ or $D_R$ in order to compute, for example, the radiated field $\psi_{\rm rad}(x)$ or the regular part of the field $\psi_R (x)$, respectively.}
\label{fig:mastersource1}
\end{figure}

In this section we demonstrate how to compute the master source through third order in $\varepsilon$. The master source can be calculated through third order in $\varepsilon$ via the Feynman diagrams given in Figure \ref{fig:mastersource1} -- these are all of the tree-level, connected diagrams that scale as $\varepsilon^3$ or lower. The solid gray line attached to a vertex represents the place to connect an ``external'' Green's function such as $D_{\rm ret} (x, z^\mu)$ if one wants to compute the radiated scalar field $\psi_{\rm rad}(x)$ or $D_R (z^\mu)$ if one wants to compute the regular part of the field on the worldline $\psi_R (z^\mu)$, for example. Compare Figure \ref{fig:mastersource1} with Figures \ref{fig:firstorderwave}-\ref{fig:thirdorderwave}.

The master source at first order is given in Figure \ref{fig:VR1} and is simply the Feynman rule for the SCO coupling to a single field perturbation, 
\begin{align}
	{\cal S}_R ^{(1)} (z^\mu) = - m c_1
\label{VR1}
\end{align}
The master source has a contribution at second order given by Figure \ref{fig:VR2}, which equals
\begin{align}
	{\cal S}_R^{(2)} (z^\mu) & = (-m c_2) \int d\tau' \, D^{-+}(z^\mu, z^{\mu'}) (-m c_1) , \\
		& = m^2 c_1 c_2 I_R (z^\mu)
\label{VR2}
\end{align}
and at third order by Figure \ref{fig:VR3},
\begin{align}
	{\cal S}_R ^{(3)} (z^\mu) = {} & (-mc_2) \int d\tau' D^{-+}(z^\mu, z^{\mu'}) (-m c_2) D^{-+}(z^{\mu'}, z^{\mu''}) (-m c_1) \nonumber \\
	& + \frac{1}{2!} ( - mc_3 ) \int d\tau' \, D^{-+}(z^\mu, z^{\mu'}) (-mc_1) \int d\tau'' \, D^{-+} (z^\mu, z^{\mu''}) (-m c_1) 
\label{VR3a} \\
	= {} & - m^3 c_1 c_2^2 \int d\tau' \, D_R (z^\mu, z^{\mu'}) I_R (z^{\mu'}) - \frac{ 1 }{ 2} m^3 c_1^2 c_3 I_R^2 (z^\mu)
\label{VR3}
\end{align}
In evaluating the diagrams in Figure \ref{fig:mastersource1} we use $D^{-+} (x,x') = D_R (x,x')$ since all of the singular integrals vanish in dimensional regularization as we have pointed out multiple times. The master source through third order is then the sum of (\ref{VR1}), (\ref{VR2}) and (\ref{VR3}),
\begin{align}
	{\cal S}_R (z^\mu) = {} & - mc_1 + m^2 c_1 c_2 I_R (z^\mu) - m^3 c_1 c_2^2 \int d\tau' \, D_R (z^\mu, z^{\mu'}) I_R (z^{\mu'}) - \frac{ 1}{2 } m^3 c_1^2 c_3 I_R^2 (z^\mu) + O(\varepsilon^4) ,
\label{VRthrough3}
\end{align}
which agrees with the expression we defined above in (\ref{source1}).
From (\ref{VRthrough3}) it is easy to obtain the radiated field $\psi_{\rm rad}(x)$ in (\ref{field5}) using (\ref{scalarrad1}), the regular part of the field on the worldline $\psi_R (z^\mu)$ in (\ref{regularfield2}) using (\ref{scalarR1}), and the regular part of the self force through third order in (\ref{renormalizedsf1}) by using (\ref{scalarR1}) and (\ref{sfR1}).

\begin{figure}
	\includegraphics[width=9cm]{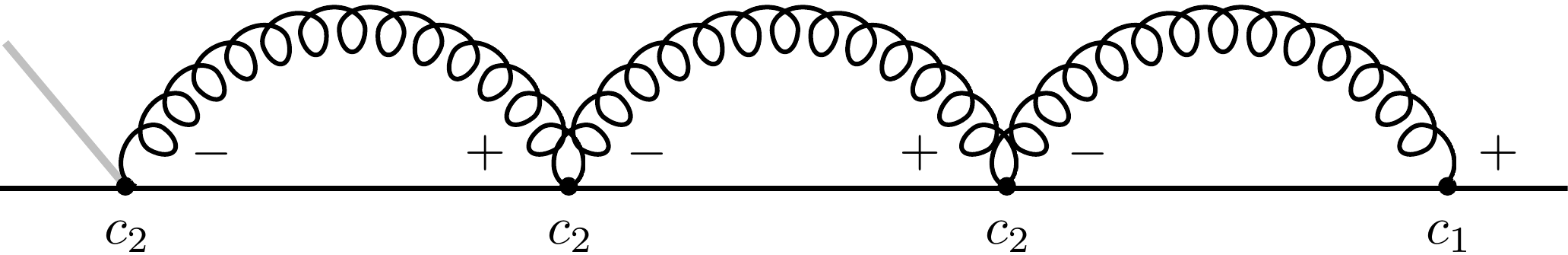} ~~
	\includegraphics[width=6.5cm]{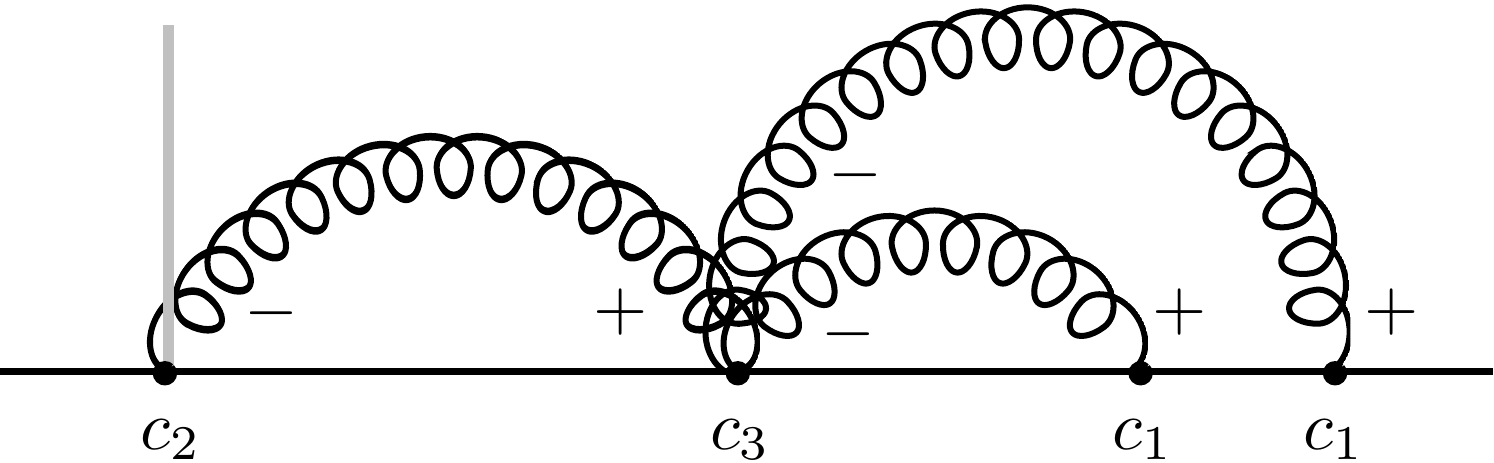} \\ ~\\
	\includegraphics[width=9cm]{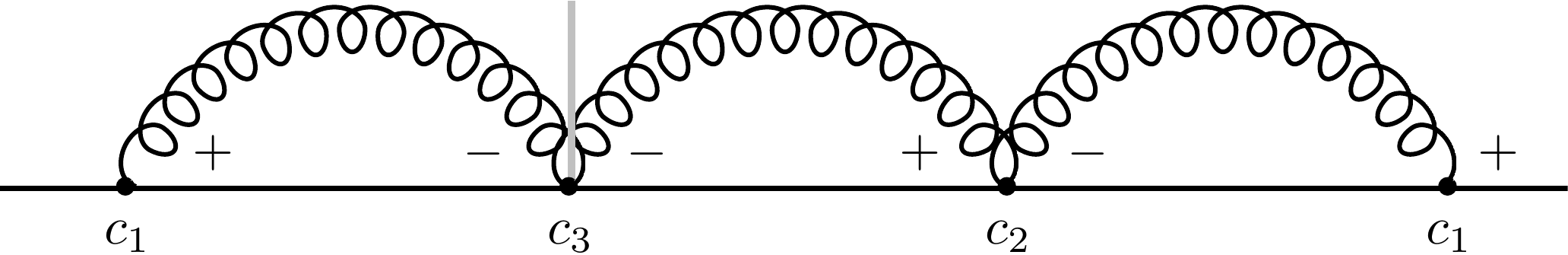} ~~
	\includegraphics[width=6.5cm]{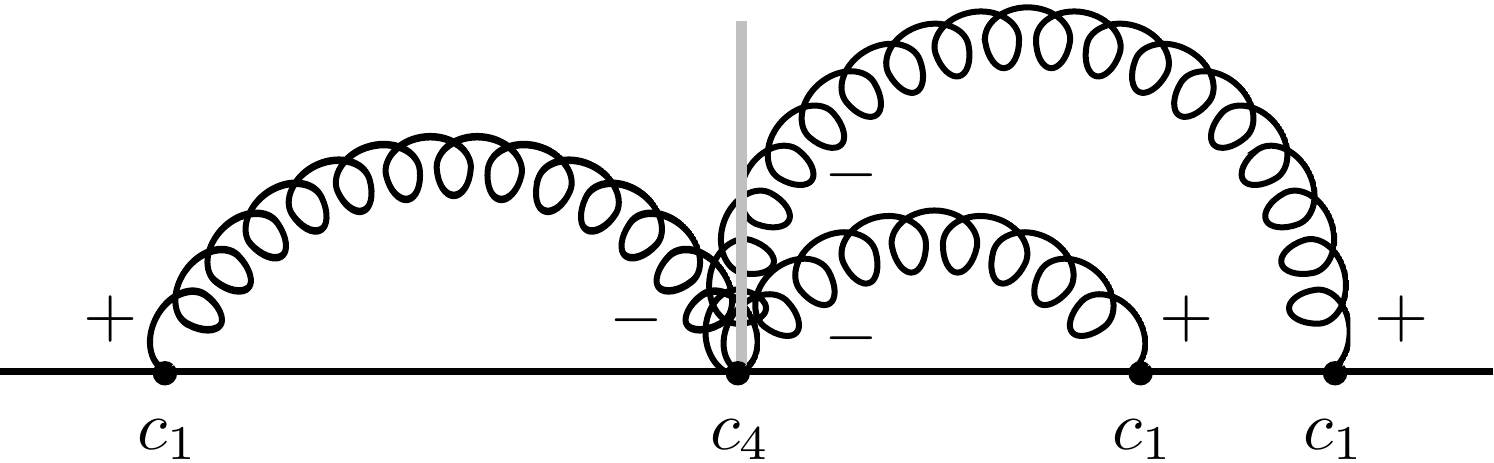} 
\caption{The diagrams contributing to the regular part of the master source ${\cal S}_R( z^\mu)$ at fourth order in $\varepsilon$. Each diagram scales as $\cR \varepsilon^4$.}
\label{fig:mastersource2}
\end{figure}

%======================================================================
\subsection{The master source at fourth order}
%======================================================================

Perturbatively computing the master source is sufficiently straightforward that one may go on to compute the fourth order corrections and higher, if desired. Figure \ref{fig:mastersource2} shows the four diagrams appearing at $O(\varepsilon^4)$. Using the Feynman rules to translate these diagrams into expressions gives
\begin{align}
	{\cal S}_R ^{(4)} (z^\mu) = {} & (-mc_2) \int d\tau' \, D^{-+}(z^\mu, z^{\mu'}) (-m c_2) \int d\tau'' \, D^{-+} (z^{\mu'}, z^{\mu''}) (-m c_2) \int d\tau''' \, D^{-+} (z^{\mu''}, z^{\mu'''}) (-m c_1) \nonumber \\
	& \frac{1}{2!} (-mc_2) \int d\tau' \, D^{-+} (z^\mu, z^{\mu'}) (-m c_3) \int d\tau'' \, D^{-+} (z^{\mu'}, z^{\mu''}) (-m c_1) \int d\tau''' \, D^{-+} (z^{\mu'}, z^{\mu'''}) (-m c_1) \nonumber \\
	& + (- m c_3) \int d\tau' \, D^{-+} (z^\mu, z^{\mu'}) (-m c_1) \int d\tau'' \, D^{-+} (z^\mu, z^{\mu''}) (-m c_2) \int d\tau''' \, D^{-+} (z^{\mu''}, z^{\mu'''}) (-m c_1) \nonumber \\
	& + \frac{1}{3!} (- m c_4) \bigg( \int d\tau' \, D^{-+} (z^\mu, z^{\mu'}) (-m c_1) \bigg)^3
\end{align}
or, again, with $D^{-+}(x,x') = D_R (x,x')$ in dimensional regularization, 
\begin{align}
	{\cal S}_R^{(4)} (z^\mu) = {} & m^4 c_1 c_2^3 \int d\tau' d\tau'' \, D_R (z^\mu, z^{\mu'}) D_R (z^{\mu'}, z^{\mu''}) I_R (z^{\mu''}) +\frac{ 1 }{ 2} m^4 c_1^2 c_2 c_3 \int d\tau' \, D_R (z^\mu, z^{\mu'} ) I_R^2 (z^{\mu'}) \nonumber \\
	&+ m^4 c_1^2 c_2 c_3 I_R (z^\mu) \int d\tau' \, D_R( z^\mu, z^{\mu'}) I_R (z^{\mu'})  + \frac{ 1 }{ 6 } m^4 c_1^3 c_4 I_R^3 (z^\mu)
\label{VR4}
\end{align}
The field radiated to a far-away observer is then simply given by (\ref{scalarrad1}) and the regular part of the field on the worldline is (\ref{scalarR1}) where the master source ${\cal S}_R$ through fourth order in $\varepsilon$ is the sum of the terms in (\ref{VR1}), (\ref{VR2}), (\ref{VR3}) and (\ref{VR4}). The contribution to the regular part of the self force at fourth order is found by reading off the appropriate terms from (\ref{sfR1}) upon substituting in (\ref{scalarR1}) in terms of the master source (we will not give the expression here).

%======================================================================
\subsection{Discussion}
%======================================================================

In the nonlinear scalar model considered here, the existence of a master source function is made possible, in part, because of the field redefinition from $\phi$ to $\psi$, which removed the self-interaction terms proportional to $\Box \phi$ from the Lagrangian thereby ensuring that
\begin{align}
	\psi(x) = \int d\tau \, D (x, z^\mu) {\cal S}_R (z^\mu)
\label{fieldmastersrc1}
\end{align}
for the appropriate propagator (e.g., retarded or $R$-part).

\begin{figure}
	\includegraphics[width=3.25cm]{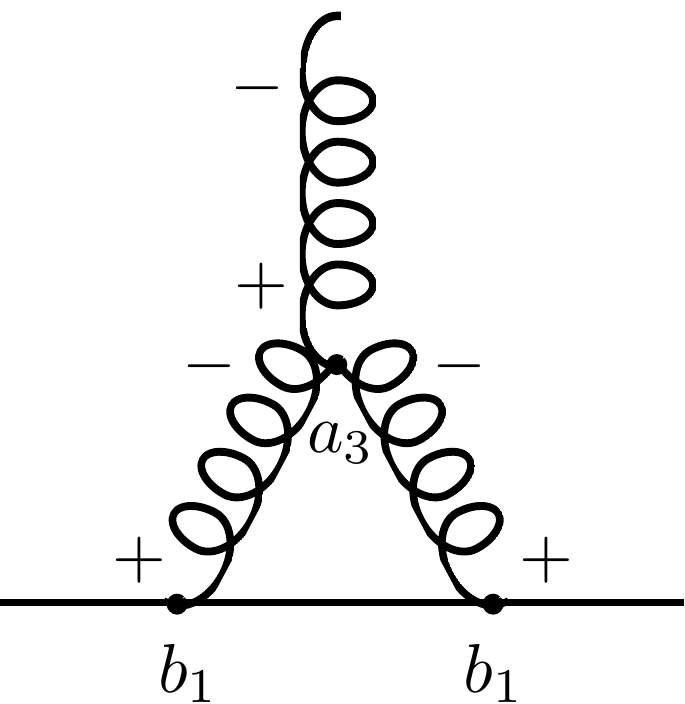}
\caption{A contribution at second order to the perturbative solution of the wave equation for $\phi(x)$.}
\label{fig:phi2}
\end{figure}

To see this, let us step back and consider the theory in the original variable $\phi(x)$. The action is given in (\ref{actionphi1}) and the corresponding wave equation in (\ref{paper1eom1}). At second order in $\varepsilon$ there is a contribution from the diagram shown in Figure \ref{fig:phi2}, which describes nonlinear self-interaction of three scalar perturbations in the bulk spacetime. The following self-interaction term in (\ref{paper1eom1})
\begin{align}
	- \frac{a_3}{3!} \int_x \phi_{,\alpha} \phi^{,\alpha} \phi = + \frac{ a_3}{12} \int_x \phi^2 \Box \phi
\label{fieldinteraction3}
\end{align}
is the quantity that accounts for the interaction of the three propagator lines in Figure \ref{fig:phi2}. The following contribution to $\phi(x)$ at second order from Figure \ref{fig:phi2} is
\begin{align}
	\phi_{(2)}(x) \supset {} &  \int d\tau d\tau' \int_y D_{\rm ret}(x, y) \bigg( \frac{a_3}{6} \bigg) \Box_y D_{\rm ret}(y, z^\mu) (-m b_1) D_{\rm ret}(y,z^{\mu'}) (-m b_1) \nonumber \\
		& + \frac{1}{2!} \int d\tau d\tau' \int_y \Box_y D_{\rm ret}(x,y) \bigg( \frac{ a_3}{6} \bigg) D_{\rm ret}(y, z^\mu) (-m b_1) D_{\rm ret}(y, z^{\mu'}) (-m b_1)
	\label{phi2a}
\end{align}
(We use $\supset$ to indicate that the right side is only one part of the full expression for $\phi_{(2)}(x)$, the other part coming from evaluating a separate diagram.) 
Using the equations of motion for the retarded and advanced Green's functions,
\begin{align}
	\Box_y D_{\rm ret} (y, x') & = - \frac{ \delta^4 (y-x') }{ g^{1/2}} \\
	\Box_{y} D_{\rm ret} (x,y) & = \Box_{y} D_{\rm adv} (y,x) = - \frac{ \delta^4 (x-y) }{ g^{1/2}}
\end{align}
we see that (\ref{phi2a}) becomes
\begin{align}
	\phi_{(2)} (x) \supset {} & - \frac{m^2 b_1^2 a_3}{6} \int d\tau d\tau' \, D_{\rm ret} (x, z^\mu) D_{R} (z^{\mu}, z^{\mu'}) - \frac{ m^2 b_1^2 a_3 }{ 12} \int d\tau d\tau' \, D_{\rm ret} (x, z^\mu) D_{\rm ret} (x, z^{\mu'})
\label{phi2b}
\end{align}
upon using dimensional regularization to evaluate the singular integral, which vanishes as usual.

The key point is that the effect of $\Box$ in (\ref{fieldinteraction3}) is to ``collapse'' a propagator line so that instead of the three propagators originally appearing in (\ref{phi2a}) one is left with only two propagators in (\ref{phi2b}). As a result, the last term of (\ref{phi2b}) contains a product of two Green's functions, each having one of its arguments evaluated at the field point $x^\mu$. Hence, the field $\phi(x)$ cannot be written at all orders in $\varepsilon$ simply as the integral of some source, as in (\ref{fieldmastersrc1}).
In fact, it is precisely the field redefinition from $\phi$ to $\psi$ in (\ref{fieldredef1}) that eliminates those self-interaction terms proportional to $\Box \phi$ (which are all of the self-interaction terms in this theory, in fact). Therefore, in the $\psi$ variable there are no contributions like the last term in (\ref{phi2b}) and one can identify a master source ${\cal S}$, as in (\ref{fieldmastersrc1}).

Using another Green's function decomposition besides Detweiler and Whiting's, such as Hadamard's decomposition into direct and tail pieces, does not yield a master source (even after changing variable from $\phi$ to $\psi$) since the tail field, while finite on the wordline, cannot simply be differentiated to yield the complete, regular self force (see, e.g., \cite{Galley:Nonlinear1}). Consequently, the self force would not be expressable in terms of a master source. It seems to us that there is a special relationship between the field redefinition in (\ref{fieldredef1}) and the Detweiler-Whiting decomposition since they both  conspire to reveal a master source ${\cal S}_R$ that generates many, if not all, physically relevant quantities for EMRIs.

%======================================================================
\section{Conclusion}
%======================================================================

In this paper we derived the scalar perturbations generated by the motion of a small compact object in a class of  nonlinear scalar models for extreme mass ratio inspirals that were developed and originally studied in Paper 1 of this series \cite{Galley:Nonlinear1}. In particular, we showed here that one can obtain sensible and finite expressions for the field perturbations generated by the SCO through at least third order in $\varepsilon = m / \cR$. We also showed that one can derive the self force corrections to the motion of the SCO from a knowledge of the regular part of the field perturbations when the latter are evaluated on the worldline. This shows, rather explicitly, that the Detweiler-Whiting scheme for computing self force corrections is valid through third order in perturbation theory. We also showed in Appendix \ref{app:allorders} that the DW scheme can be applied at any order in perturbation theory since the regular self force is found, at all orders in $\varepsilon$, to be given by
\begin{align}
	F_R^\mu (\tau) = - m (a^\mu + P^{\mu\nu} \nabla_\nu) \sum_{n=1}^\infty \frac{ c_n}{n!} \psi_R^n (z^\mu)
\end{align}

The divergent integrals that we encountered in the course of the calculations here are subtracted by exactly the same counter terms found in Paper 1. This explicitly demonstrates the internal consistency of our regularization and renormalization program. If using dimensional regularization to regularize the integrals, then all singular contributions vanish since the nonlinear scalar model considered here cannot generate any logarithmically divergent integrals. This property greatly streamlines the derivation of higher-order corrections of the field and self force (see Section \ref{sec:efficient}). 

For a specific choice of parameters in our nonlinear scalar model we recovered the second-order perturbative solution for the field $\phi(x)$ that was first derived using very different methods by Rosenthal \cite{Rosenthal:CQG22}. The method of counter terms used in this series of papers yields exactly the same finite part as the series of steps carefully developed in \cite{Rosenthal:CQG22}. In addition, using dimensional regularization to evaluate the singular integrals makes the calculations even more efficient as the power-divergent, singular integrals subsequently vanish. We expect that our regularization prescription will be most beneficial, if not crucial, for deriving higher-order self force expressions in the gravitational description of EMRIs.

All of the results contained in this paper and in Paper 1 can be derived from one quantity, which is the regular part of the master source ${\cal S}_R(z^\mu)$. The master source can be integrated with the retarded Green's function to generate the (scalar) waveform that would be measured by an observer far away, it can be integrated with the regular part of the retarded Green's function $D_R(x,z^\mu)$ to yield the regular part of the field on the worldline and, consequently, the regular part of the self force on the SCO. It appears that most, if not all, relevant physical quantities can be derived by calculating the master source up to the desired order in $\varepsilon$. We also showed how the master source can be calculated in perturbation theory using Feynman diagrams and proceeded to demonstrate this with a derivation of the master source through \emph{fourth} order in $\varepsilon$.

We conjecture that a master source ${\cal S}^{\alpha \beta}_R (x^\mu)$ exists for the gravitational description of EMRIs. Preliminary calculations suggest that a change of variable from $h_{\alpha \beta}$ to another field $H_{\alpha \beta}$ can be used to eliminate all cubic self-interaction terms in the action that are proportional to $\Box (h_{\alpha \beta} - g_{\alpha \beta} h^\gamma _\gamma /2)$, just as in the nonlinear scalar model. It seems likely that such a change of variable can be constructed perturbatively to remove higher order interaction terms proportional to $\Box (h_{\alpha \beta} - g_{\alpha \beta} h^\gamma _\gamma /2)$. Therefore, using this field redefinition together with the Detweiler-Whiting decomposition of Green's functions into regular and singular parts, we expect that the gravitational perturbations generated by the SCO in the new variable $H_{\alpha \beta}$ may have a representation, \emph{at all orders in $\varepsilon$}, of the form 
\begin{align}
	H_{\alpha \beta} (x) = \int_{x'} D_{\alpha \beta \gamma' \delta'} (x, x') {\cal S}^{\gamma' \delta'}_R (x')
\end{align}
and expressed in terms of a regular, master source function ${\cal S}_R^{\alpha \beta} (x^\mu)$, at least in the Lorenz gauge for trace-reversed metric perturbations in a vacuum, background spacetime. If true, one will be able to derive the gravitational waveform measured by an observer, the regular part of the metric perturbations $H^R _{\alpha \beta}(z^\mu)$ (and its derivative) on the worldline of the SCO, the regular part of the self force, gauge-invariant variables when restricting to conservative dynamics, etc., by computing the perturbative expression for essentially one quantity, the master source, to the desired order in $\varepsilon$.

%================
\acknowledgments
%================

We thank Tanja Hinderer for providing valuable feedback and comments on a previous draft.
This work was supported in part by an appointment to the NASA Postdoctoral Program at the Jet Propulsion Laboratory adminstered by Oak Ridge Associated Universities through a contract with NASA. Copyright 2011. All rights reserved.

\appendix

%======================================================================
\section{Feynman rules for computing scalar perturbations}
\label{app:feynman}
%======================================================================

The Feynman rules are adapted (and can be derived) from Paper 1 for the specific purpose of computing scalar perturbations. Since we will not be computing an effective action in this paper then we do not need to consider two sets of histories for the worldline and field, etc., as outlined in Paper 1.

The Feynman rules used in this paper are as follows:
\begin{itemize}

\item{For each worldline vertex represented by a dark circle and labeled by $c_n$ write down a factor of $(-m c_n)$. These are the worldline vertices.}

\item{For each curly line write down a factor of $D^{-+}( z^\mu , z^{\mu'}) \equiv D_{\rm ret} (z^\mu , z^{\mu'})$ connecting worldline vertices labeled by  ``$-$'' and ``$+$'' at proper times $\tau$ and $\tau'$, respectively. These are the propagators.}

\item{Integrate over all proper times.}

\item{Divide by the appropriate symmetry factor.}

\end{itemize}
As an example, apply these rules to the diagram on the right in Figure \ref{fig:thirdorderwave}. The Feynman rules listed above imply that the diagram equals
\begin{align}
	\frac{1}{2!} \int d\tau d\tau' d\tau'' \, D^{-+}(x, z^\mu) (-m c_3) D^{-+} (z^\mu, z^{\mu'}) (-m c_1) D^{-+}(z^\mu, z^{\mu''}) (-m c_1)
\label{example1}
\end{align}
The first factor is the reciprocal of the symmetry factor, which is $2!$ because there are two ways to connect the ends of the two propagators emanating from the $c_3$ vertex to the two $c_1$ vertices. The first propagator $D^{-+}(x, z^\mu)$ in (\ref{example1})  is the retarded propagator connecting the $c_3$ vertex at $z^\mu(\tau)$ to the field point at $x^\mu$. The factor of $(-m c_3)$ is the worldline vertex where the three propagators all connect to the worldline. There are two propagators stemming from the $c_3$ vertex and connecting from $z^\mu(\tau)$ back to the worldline at proper times $\tau'$ and $\tau''$ through the two $(-m c_1)$ worldline vertices. This example shows how one can use the Feynman rules listed in this Appendix to construct any such diagram for computing scalar perturbations at any order in perturbation theory. Power counting (\ref{example1}) using the scaling laws in Section \ref{sec:wavesineft} reveals that the diagram on the right in Figure \ref{fig:thirdorderwave} scales as (see (\ref{example1}))
\begin{align}
	\sim (\cR)^3 (\cR^{-2}) (m) (\cR^{-2}) (m) (\cR^{-2}) (m) = \frac{ m^3 }{ \cR^3 } = \varepsilon^3
\end{align}
and is thus a contribution at third order in perturbation theory.

%======================================================================
\section{Detweiler-Whiting scheme at all orders in perturbation theory}
\label{app:allorders}
%======================================================================

In this Appendix, we prove that a regular expression for the master source can be found through any order in perturbation theory. We will use dimensional regularization to evaluate the singular integral $\Lambda$, in which case $\Lambda = 0$.
From the form of the action in (\ref{actionpsi1}) it is clear that all interactions, by construction from the field redefinition, are confined to be on the worldline. Since solutions involving integrals of propagators evaluated at the same point (e.g., $D(z^\mu(\tau) , z^\mu(\tau))$) cannot be generated in a classical theory then every singular contribution to the master source contains divergent integrals from the set
\begin{align}
	  \bigg\{ \int d\tau_2 \cdots d\tau_{n} \, D_{\rm ret} (z^{\mu_1}, z^{\mu_2}  ) \cdots D_{\rm ret} (z^{\mu_{n-1} } , z^{\mu_n}  ) \bigg\} _{n=1} ^N \in {\cal S} _{(N)} (z^{\mu_1})
\label{genintegral1}
\end{align}
where $z^{\mu_i} \equiv z^\mu (\tau_i)$ for some positive integer $i$. Let the $n^{\rm th}$ member of this set be called $J_n (z^{\mu_1})$. For example, ${\cal S}_{(3)}$ in (\ref{VR3a}) contains a $J_2 (z^{\mu_1})$ integral and two $J_1 (z^{\mu_1})$ integrals.

Consider $J_n (z^{\mu_1})$, any member in the set of (\ref{genintegral1}), and first regularize the $\tau_n$ integral. In the DW scheme, one writes the retarded propagator as
\begin{align}
	D_{\rm ret} (x, x') = D_S (x, x') + D_R (x,x')
\end{align}
so that
\begin{align}
	J_n (z^{\mu_1}) = \int d\tau_2 \cdots d\tau_{n-1}  \bigg( \prod _{k=1}^{n-2} D_{\rm ret} (z^{\mu_k}, z^{\mu_{k+1}} ) \bigg) \int d\tau_n \, \Big[ D_R (z^{\mu_{n-1}}, z^{\mu_n} ) + D_S (z^{\mu_{n-1}}, z^{\mu_n} ) \Big]
\end{align}
From (\ref{singularintegral1}) we know that the $\tau_n$ integral is just $\Lambda/ (4\pi) + I_R (z^{\mu_{n-1}} ) = I_R(z^{\mu_{n-1}})$ thereby yielding
\begin{align}
	J_n(z^{\mu_1}) = {} & \int d\tau_2 \cdots d\tau_{n-2}  \bigg( \prod _{k=1}^{n-3} D_{\rm ret} (z^{\mu_k}, z^{\mu_{k+1}} ) \bigg)  \int d\tau_{n-1} \, \bigg[ D_R (z^{\mu_{n-2}}, z^{\mu_{n-1}} ) + D_S (z^{\mu_{n-2}}, z^{\mu_{n-1}} ) \bigg]  I_R (z^{\mu_{n-1}} ) 
\label{genintegral2}
\end{align}
The $\tau_{n-1}$ integral is then
\begin{align}
	\int d\tau_{n-1} D_R(z^{\mu_{n-2}}, z^{\mu_{n-1}}) I_R (z^{\mu_{n-1}}) + \int d\tau_{n-1} D_S(z^{\mu_{n-2}}, z^{\mu_{n-1}} ) I_R (z^{\mu_{n-1}}) 
\label{genintegral4}
\end{align}
The second term involves a proper time integral over $D_S$ and a regular function of $\tau_{n-1}$. Since the latter is regular we may expand it in a Taylor series for $s_{n-1} \equiv \tau_{n-1} - \tau_{n-2}$ near zero
\begin{align}
	I_R (z^{\mu_{n-1}}) = I_R (z^{\mu_{n-2}}) + s_{n-1} \dot{I}_R (z^{\mu_{n-2}}) + O(s_{n-1}^2)
\end{align}
where a dot represents $d/d\tau_{n-2}$. Using (\ref{lambda1}) and the above expression, the second term in (\ref{genintegral4}) becomes
\begin{align}
	 \int_{-\infty}^\infty \!\!\! ds_{n-1} \, \frac{ 1}{8\pi} \frac{\delta (s_{n-1}) }{ | s_{n-1} | } \bigg( I_R (z^{\mu_{n-2}}) + s_{n-1} \dot{I}_R (z^{\mu_{n-2}}) + O(s_{n-1}^2) \bigg)
\label{genintegral6}
\end{align}
Since
\begin{align}
	\int_{-\infty}^{\infty} ds_{n-1} \frac{ s_{n-1}^p }{ | s_{n-1} | } \delta (s _{n-1}) = 0
\end{align}
for $p\ge 1$
then it follows that (\ref{genintegral6}) equals
\begin{align}
	\int d\tau_{n-1} D_S(z^{\mu_{n-2}}, z^{\mu_{n-1}} ) I_R (z^{\mu_{n-1}}) = \frac{ \Lambda }{ 4\pi} I_R (z^{\mu_{n-2}}) = 0
\end{align}
Thus, (\ref{genintegral2}) is
\begin{align}
	J_n(z^{\mu_1}) = {} & \int d\tau_2 \cdots d\tau_{n-2}  \bigg( \prod _{k=1}^{n-3} D_{\rm ret} (z^{\mu_k}, z^{\mu_{k+1}} ) \bigg) \int d\tau_{n-1} \,  D_R (z^{\mu_{n-2}}, z^{\mu_{n-1}} )   I_R (z^{\mu_{n-1}} )
\end{align}
The remaining proper time integrals are evaluated in like manner. In particular, by induction it follows that
\begin{align}
	J_n(z^{\mu_1}) = \int d\tau_2 \cdots d\tau_{n}  \bigg( \prod _{k=1}^{n-1} D_{R} (z^{\mu_k}, z^{\mu_{k+1}} ) \bigg)   
\end{align} 
and the regular master source is constructed from integrals in the set appearing in the right side of (\ref{genintegral1}) with $D_{\rm ret}$ replaced by the regular part, $D_R$. Hence, one can always find the regular part of the master source ${\cal S}_R (z^{\mu_1})$ to any order in perturbation theory.

Now for the self force. Since we have shown that the regular part of the master source can always be constructed at any order in perturbation theory then the field is simply
\begin{align}
	\psi(x) = \psi_R (x) = \int d\tau' \, D_R (x, z^{\mu'}) {\cal S}_R(z^{\mu'})
\label{appfield1}
\end{align}
which is regular on the worldline when $x^\mu = z^\mu(\tau)$. The self force also depends on the derivative of $\psi$ on the worldline. Using calculations presented in Paper 1 it is straightforward to show that
\begin{align}
	P^{\mu\nu} \nabla_\nu \psi(z^\mu) & = - \frac{ \Lambda }{ 4 \pi } \frac{ a^\mu }{ 2} {\cal S}_R(z^\mu) + P^{\mu\nu} \nabla_\nu \psi_R (z^\mu ) \\
		& = P^{\mu\nu} \nabla_\nu \psi_R (z^\mu)
	\label{appderfield1}
\end{align}
Rewriting the worldline equations of motion in (\ref{eom2}) as
\begin{align}
	m a^\mu = - m ( a^\mu + P^{\mu\nu} \nabla_\nu ) C( \psi(z^\mu)) = F^\mu (\tau) 
\end{align}
it follows from (\ref{appfield1}) and (\ref{appderfield1}) that the self force on the small compact object is 
\begin{align}
	F^\mu (\tau) = - m ( a^\mu + P^{\mu\nu} \nabla_\nu ) C ( \psi_R (z^\mu) ) = F^\mu_R (\tau)
\end{align}
Therefore, the Detweiler-Whiting scheme for calculating the self force is valid at any order in perturbation theory, at least in this class of nonlinear scalar theories.  
Of course, $\psi_R (z^\mu)$ must be computed perturbatively and we performed the calculation explicitly in this paper through third order in $\varepsilon$ and the master source through fourth order. Indeed, the EFT approach provides a systematic way of computing $\psi_R$ to any order in perturbation theory using Feynman diagrams.

\input{Bibliography}

\end{document}

%% file: Bibliography.tex
\bibliographystyle{physrev}
\bibliography{gw_bib}

\setlength{\parskip}{1em}